\documentclass[a4paper,11pt]{article}
\pdfoutput=1 
\usepackage{jcappub} 
\usepackage[T1]{fontenc}
\usepackage[table]{xcolor}
\usepackage{epsfig}
\usepackage{colortbl}
\usepackage[T1]{fontenc}
\usepackage{tikz-feynman}
\usepackage{hepnames}
\usepackage{adjustbox}
\usepackage{textcomp}
\usepackage{slashed}
\usepackage{amsmath}
\usepackage{bbm}
\usepackage{multirow}
\usepackage[normalem]{ulem}

\newcommand{\gss}{g_{\star s}}
\newcommand{\Trh}{T_\text{rh}}
\newcommand{\arh}{a_\text{rh}}
\newcommand{\Tmax}{T_\text{max}}

\newcommand{\rR}{\rho_R}
\newcommand{\rp}{\rho_\phi}

\newcommand{\Gp}{\Gamma_\phi}

\newcommand{\mueff}{\mu_\text{eff}}
\newcommand{\yeff}{y_\text{eff}}
\newcommand{\ndm}{n_{\rm DM}}
\newcommand{\Ndm}{N_{\rm DM}}
\newcommand{\mdm}{m_{\rm DM}}
\newcommand{\adm}{a_{\rm DM}}
\newcommand{\lNP}{\Lambda_{\rm NP}}

\DeclareUnicodeCharacter{2212}{-}
\newcommand{\bea}{\begin{eqnarray}}
\newcommand{\eea}{\end{eqnarray}}
\usepackage{xcolor}

\usepackage[font=small]{caption}
\usepackage[font={small},labelfont={bf}]{caption}
\usepackage{url}
\definecolor{MyDarkBlue}{rgb}{0.1, 0.1, 0.8}
\definecolor{SBlue}{rgb}{0.2, 0.4, 0.7} 
\definecolor{MyLightBlue}{rgb}{0.22,0.51,0.9}
\definecolor{MyGreen}{rgb}{0.0, 0.5, 0.0}
\definecolor{BrickRed}{rgb}{0.8, 0.25, 0.33}
\usepackage{hyperref}
\title{Lepton collider as a window to reheating via freezing in dark matter detection. Part II}
\author[a]{Basabendu Barman,}
\author[b]{Subhaditya Bhattacharya,}
\author[b]{Sahabub Jahedi,}
\author[b]{Dipankar Pradhan,}
\author[b]{and Abhik Sarkar}
\affiliation[a]{Department of Physics, School of Engineering and Sciences, SRM University-AP, Amaravati 522240, India}
\affiliation[b]{Department of Physics, Indian Institute of Technology Guwahati,\\	North Guwahati, Assam-781039, India,}
\emailAdd{basabendu.b@srmap.edu.in}
\emailAdd{subhab@iitg.ac.in}
\emailAdd{sahabub@alumni.iitg.ac.in}
\emailAdd{d.pradhan@iitg.ac.in}
\emailAdd{sarkar.abhik@iitg.ac.in}
\abstract{Dark matter (DM) genesis via Ultraviolet (UV) freeze-in embeds the seed of reheating temperature and dynamics in its relic density. Thus, discovery of such a DM candidate can possibly open the window for post-inflationary dynamics. However, there are several challenges in this exercise, as freezing-in DM possesses feeble interaction with the visible sector and therefore very low production cross-section at the collider. We show that mono-photon (and dilepton) signal at the ILC, arising from DM effective operators connected to the SM field strength tensors, can still warrant a signal discovery. We study both the scalar and fermionic DM production during reheating via UV freeze-in, when the inflaton oscillates at the bottom of a general monomial potential. Interestingly, we see, right DM abundance can be achieved only in the case of bosonic reheating scenario, satisfying bounds from big bang nucleosynthesis (BBN). This provides a unique correlation between collider signal and the post-inflationary dynamics of the Universe within single-field inflationary models.
}
\begin{document} 
%%%%%%%%%%%%%
%%%%%%%%%%%%%
\makeatletter
\gdef\@fpheader{}
\makeatother
%%%%%%%%%%%%%
\maketitle
%%%%%%%%%%%%%%%
\section{Introduction}
\label{sec:intro}
%%%%%%%%%%%%%%%%
Over the past few decades, our comprehension of cosmic history has undergone a remarkable transformation. Precise measurements of temperature anisotropies in the cosmic microwave background (CMB) have unveiled a Universe that is homogeneous and isotropic on large scales. These observations provide compelling evidence for an inflationary phase in the early Universe. Furthermore, the relative abundances of light elements align exceptionally well with theoretical expectations, reinforcing the predictions of big-bang nucleosynthesis (BBN). These predictions, grounded in the well-understood physics of nuclear reactions, point to a hot and dense Universe in local thermal equilibrium during later stages. However, bridging these two extraordinary epochs—inflation and BBN—remains a formidable challenge. The energy scale of inflation is as high as $\sim 10^{16}$ GeV, while that of BBN is about 4 MeV. This immense range of energy (and therefore, time) scale is still poorly understood and remains largely unconstrained by observations. Nevertheless, it is necessary to have a closer look into the period between inflation and BBN and the dynamics of reheating as it holds the key to the state of the Universe that we observe today. Importantly, the process of reheating not only explains the cosmic origin of the matter that constitutes our physical world but also accounts for the production of relics beyond the Standard Model (SM), such as, dark matter (DM). Particle colliders and accelerator-based experiments possess the profound capability of replicating some of the Universe's earliest epochs. Colliders, such as the Large Hadron Collider (LHC), have demonstrably established their exceptional efficacy in elucidating the intricate mechanisms underlying the electroweak symmetry breaking (EWSB), a phenomenon that transpired at an energy scale on the order of $\sim\mathcal{O}(10^2)$ GeV, as well as the QCD phase transition, which occurred at an energy scale of approximately $\sim\mathcal{O}(10^2)$ MeV. Thus, colliders emerge as promising instruments for the study of early Universe dynamics within the controlled environment of laboratories.

Stringent observational constraints (see, e.g., Refs.~\cite{Roszkowski:2017nbc, Arcadi:2017kky, Chang:2019xva, Darme:2019wpd}) on the typical weakly interacting massive particle (WIMP) parameter space have prompted the exploration of alternative DM production mechanisms. One such alternative is the idea of feebly interacting massive particles (FIMPs). FIMPs can be generated in the early Universe through the decay or annihilation of particles in the visible sector. When the temperature of the SM bath drops below the typical interaction mass scale—defined as the maximum of the DM and mediator masses—the DM production process becomes Boltzmann suppressed, leading to a constant comoving DM number density, thereby achieving freeze-in~\cite{McDonald:2001vt, Hall:2009bx, Bernal:2017kxu}. The FIMP paradigm requires extremely suppressed interaction rates between the dark and visible sectors, in order to ensure non-thermal production of DM. This is achievable either through tiny couplings in its infrared (IR) avatar or through non-renormalizable operators suppressed by a new physics (NP) scale, in case of ultraviolet (UV) freeze-in~\cite{Hall:2009bx,Elahi:2014fsa}. The UV scenario is particularly intriguing because the DM yield becomes sensitive to the highest temperature $\Tmax$ reached by the SM plasma~\cite{Giudice:2000ex,Bernal:2019mhf,Bernal:2020qyu,Barman:2020plp,Garcia:2020eof,Garcia:2020wiy,Ahmed:2021try,Mambrini:2021zpp,Kaneta:2021pyx,Ahmed:2021fvt,Barman:2022tzk,Haque:2023yra,Becker:2023tvd,Bernal:2024ykj,Barman:2024ujh}, which is determined by the inflaton dynamics. 

Building on these motivations, this work explores the potential to illuminate the reheating period through collider searches for DM, specifically by examining their production during the post-inflationary epoch. In order to establish a direct correspondence between the collider signature of DM and the early Universe dynamics, we have resorted to DM production via UV freeze-in since in this case, the DM yield becomes important around $\Tmax$. We have considered all of the DM relic was produced from the thermal bath during reheating, while the bath itself is assumed to be sourced by the perturbative decay of the inflaton field either into a pair of bosons or into a pair of fermions. Consequently, any potential DM signature, such as mono-$X$ (where $X$ can be any SM particle observed at colliders, for example, a photon) plus missing energy, could be directly correlated with (a) the maximum temperature reached during reheating and (b) the shape of the potential during this phase. Our analysis is centred on lepton colliders—due to their relatively low SM background. We highlight the case of photophilic DM, which allows for the detection of mono-photon events emerging from the vertex, allowing a nice signal background separation. LHC, plagued with huge backgrounds for mono-photon signal, may find it difficult to probe such kind of events even with high luminosity run. The same connection can however be established for some other kinds of operators. This work is an extension of Ref.~\cite{Barman:2024nhr}, where we considered instantaneous reheating, and therefore overlooked the evolution of the thermal bath during the entire stage of reheating. The present work, therefore, is a major improvement over Ref.~\cite{Barman:2024nhr}, where, as we will show, the allowed parameter space for the DM opens up to a large extent and becomes accessible for collider probe on simply considering non-instantaneous decay of the inflaton field.

The paper is organized as follows. The underlying particle physics model is discussed in Sec.~\ref{sec:model}. In Sec.~\ref{sec:reheating} we have detailed the post-inflationary dynamics of the inflaton, together with DM production. We then discuss the collider analysis in Sec.~\ref{sec:collider}, where we elaborate on the connection between collider signal and reheating. Finally, we conclude in Sec.~\ref{sec:concl}.
%%%%%%%%%%%%%%%
\section{Particle physics model for DM-SM interactions}
\label{sec:model}
%%%%%%%%%%%%%%%
As advocated in the beginning, we are interested in the production of DM via UV freeze-in. UV freeze-in occurs mostly in cases where the accessible bath temperature is way higher than the dark sector particle masses, and therefore best suited for non-renormalisatble operators producing DM via scattering of bath particles. We shall therefore focus on non-renormalizable interactions involving a pair of SM and a pair of DM fields. Now, at colliders, because of the absence of electromagnetic interactions, the DM evades detection, making it challenging to observe. However, the visible particles produced either from initial state radiation (ISR) or in association with DM can provide crucial insights by revealing momentum or energy imbalance. These imbalances are key indicators in DM searches. A well-known approach in this context is the search for events involving mono-$X$ signatures, where $X$ can represent a photon ($\gamma$), jet ($j$), $Z$ boson, or Higgs boson, accompanied by missing energy ($\slashed{E}$)  \footnote{Mono-$X$ final states have been extensively studied in the context of lepton colliders~\cite{Fox:2011fx, Yu:2013aca, Essig:2013vha, Kadota:2014mea, Yu:2014ula, Freitas:2014jla, Dutta:2017ljq, Choudhury:2019sxt, Horigome:2021qof, Barman:2021hhg, Kundu:2021cmo, Bhattacharya:2022qck, Ge:2023wye, Ma:2022cto, Singh:2024wdn, Borah:2024twm}.}. Among these, the mono-$\gamma$ final state is particularly interesting as we explain later. A notable challenge in this search arises from the fact that the ISR photon events have a missing energy distribution similar to that of the substantial SM neutrino background ($\nu\bar{\nu}\gamma$). This complicates the separation of signal from background, as we shall demonstrate in Sec.~\ref{sec:collider}. As a consequence, we concentrate on the production of associated photons, which we refer to as the {\em natural} mono-$\gamma$ signal. This basically means we want the photon to be produced along with the DM pair from the same vertex at the collider, not as an ISR of electron(positron) beams.

Such a constraint restricts us to look into a rather small set of operators relevant for mono-photon production. We construct dimension-six and seven effective operators with real spin-0 boson $\Phi$, massive spin-1 gauge boson $X_\alpha$ and Dirac-type spin-1/2 fermion $\chi$ as DM candidates, that can couple to the weak gauge boson field strength tensors to potentially provide natural mono-photon signal\footnote{Such operators have been explored also in the context of freeze-out~\cite{Nelson:2013pqa,Arina:2020mxo,Kavanagh:2018xeh}.},
%%%%%%%%%%%%%%%%%%%%%%%%%%%
\begin{align}     
\begin{split}
& \mathcal{O}_{2}^{s}:~\dfrac{c_{\Phi}}{\Lambda^2_{\rm NP}}(B_{\mu\nu}B^{\mu\nu}+W_{\mu\nu}^i\,W^{i\mu\nu})\,\Phi^2\,,
%\label{eq:os2}
\\& \mathcal{O}_{3}^{f}:~\dfrac{c_{\chi}}{\Lambda^3_{\rm NP}}(B_{\mu\nu}B^{\mu\nu}+W_{\mu\nu}^i\,W^{i\mu\nu})\,(\overline{\chi}\chi)\,,
%\label{eq:of3}
\\& \mathcal{O}_{2}^V:~\dfrac{c_V}{\Lambda^2_{\rm NP}}(B_{\mu\nu}B^{\mu\nu}+W_{\mu\nu}W^{\mu\nu})\,X_\alpha\,X^\alpha\,,
%\label{eq:ova}
\\& \mathcal{O}_{4}^{VV}:~\dfrac{c_{VV}^{}}{\Lambda^4_{\rm NP}}(B_{\mu\nu}B^{\mu\nu}+W_{\mu\nu}W^{\mu\nu})\,X_{\alpha\beta}\,X^{\alpha\beta}\,.
\label{eq:eft-ops}
\end{split}
\end{align}
%%%%%%%%%%%%%%%%%%%%%%%%%%%
Here $W_{\mu \nu}^i$ $(i=1,\,2,\,3)$ and $B_{\mu \nu}$ are the electroweak field strength tensors corresponding to $SU(2)_L$ and $U(1)_Y$ respectively, $X_{\alpha\beta}$ represents field strength tensor for the vector boson DM, $\Lambda_{\rm NP}$ defines the scale of NP that generates DM-SM effective interaction, and $c_i$'s are the dimensionless Wilson-coefficients (WCs). Unless otherwise explicitly mentioned, we consider all the WCs to be unity. Note that as all these operators are loop generated when the NP is weakly coupled, they have an additional suppression of $1/(16 \pi^2)$, therefore requiring the NP scale to be appropriately rescaled. The presence of DM fields in pair implies $\mathbb{Z}_2$-symmetry that ensures absolute stability of the DM. \textcolor{black}{For fermionic DM, there also exists a dimension-five operator, $\mathcal{O}^f_1 \sim B_{\mu \nu} (\bar{\chi} \sigma^{\mu \nu} \chi)$, that can mimic a natural mono-photon-like signal via final state radiation (FSR). However, the phenomenology of such an operator is strongly constrained by DM relic abundance, pushing the corresponding NP scale beyond the reach of the collider set-up considered here.} In the rest of the analysis, we will typically concentrate on the spin-0 and spin-1/2 scenarios. All the relevant annihilation cross-sections for DM production is reported in Appendix~\ref{sec:cs}. 
%%%%%%%%%%%%%%%%%%%%%%%%%
\section{Post-inflationary inflaton dynamics}
\label{sec:reheating}
%%%%%%%%%%%%%%%%%%%%%%%%%%%%
In this section, we shall elaborate on the freeze-in production of DM during reheating. We therefore begin by discussing the post-inflationary dynamics of the inflaton that results in generation of the radiation bath, that in turn acts as a source for DM. 

We consider the post-inflationary oscillation of the inflaton $\phi$ in a monomial potential
\begin{equation}\label{eq:inf-pot}
V(\phi) = \lambda\, \frac{\phi^n}{\Lambda^{n - 4}}\,,
\end{equation}
where $\lambda$ is a dimensionless coupling and $\Lambda\lesssim 10^{16}$ GeV is the scale of inflation, which is bounded from above from CMB measurements of the inflationary parameters~\cite{Planck:2018jri}\footnote{The potential in Eq.~\eqref{eq:inf-pot} can naturally arise in a number of inflationary scenarios, for example, the $\alpha$-attractor T- or E-models~\cite{Kallosh:2013hoa, Kallosh:2013yoa, Kallosh:2013maa}, or the Starobinsky model~\cite{Starobinsky:1980te, Starobinsky:1981vz, Starobinsky:1983zz, Kofman:1985aw}. Given a particular inflationary model, $\lambda$ can be fixed from CMB measurements of the inflationary observables.}. Also note that the NP that generates $\Lambda$ is 
quite different from that of $\Lambda_{\rm NP}$ characterizing DM-SM interactions, and thus can be widely different. The effective mass $m_\phi(a)$ for the inflaton can be obtained from the second derivative of Eq.~\eqref{eq:inf-pot}, which reads
\begin{equation}\label{eq:inf-mass1}
m_\phi(a)^2 \equiv \frac{d^2V}{d\phi^2} = n\, (n - 1)\, \lambda\, \frac{\phi^{n - 2}}{\Lambda^{n - 4}}
\simeq n\, (n-1)\, \lambda^\frac{2}{n}\, \Lambda^\frac{2\, (4 - n)}{n} \rp(a)^{\frac{n-2}{n}}\,.
\end{equation}
Note, for $n \neq 2$, $m_\phi$ has a field dependence that, in turn, would lead to a time-dependent inflaton decay rate.

The equation of motion for the oscillating inflaton condensate reads~\cite{Turner:1983he}
\begin{equation} \label{eq:eom0}
\ddot\phi + (3\, H + \Gp)\, \dot\phi + V'(\phi) = 0\,,
\end{equation} 
where $H$ denotes the Hubble expansion rate, $\Gp$ is the inflaton decay rate, which we will elaborate on in a moment. Here the dots represent derivatives with respect to the time $t$, while primes are derivatives with respect to the field. The evolution of the inflaton energy density $\rp \equiv \frac12\, \dot\phi^2+ V(\phi)$ can be tracked with the following Boltzmann equation (BEQ)
\begin{equation} \label{eq:drhodt}
\frac{d\rp}{dt} + \frac{6\, n}{2 + n}\, H\, \rp = - \frac{2\, n}{2 + n}\, \Gp\, \rp\,,
\end{equation}
where $H=\sqrt{\left(\rp+\rR\right)/(3\,M_P^2)}$ is the Huuble rate. It is important to note here that the inflaton equation of state is parametrized as $w \equiv p_\phi/\rp = (n - 2) / (n + 2)$~\cite{Turner:1983he}, where $p_\phi \equiv \frac12\, \dot\phi^2 - V(\phi)$ is the pressure. During reheating $a_I \ll a \ll \arh$, where $a$ is the scale factor of the Universe, the term associated with expansion, $H\, \rp$ typically dominates over the interaction term $\Gp\, \rp$. Then it is possible to solve Eq.~\eqref{eq:drhodt} analytically, leading to
\begin{equation} \label{eq:rpsol}
\rp(a) \simeq \rp (\arh) \left(\frac{\arh}{a}\right)^\frac{6\, n}{2 + n}.
\end{equation}
Here, $a_I$ and $\arh$ correspond to the scale factor at the end of inflation and at the end of reheating, respectively. Since the Hubble rate during reheating is dominated by the inflaton energy density, one can write from Eq.~\eqref{eq:rpsol}
\begin{equation} \label{eq:Hubble}
H(a) \simeq H(\arh) \times
\begin{cases}
\left(\frac{\arh}{a}\right)^\frac{3\, n}{n + 2} &\text{ for } a \leq \arh\,,\\[10pt]
\left(\frac{\arh}{a}\right)^2 &\text{ for } \arh \leq a\,.
\end{cases}
\end{equation}
The evolution of the SM radiation energy density $\rR$, on the other hand, is governed by the Boltzmann equation of the form~\cite{Garcia:2020wiy}
\begin{equation} \label{eq:rR}
\frac{d\rR}{dt} + 4\, H\, \rR = + \frac{2\, n}{2 + n}\, \Gp\, \rp\,.
\end{equation}
The end of reheating is defined by
\begin{align}
\rR(\arh) = \rp(\arh) = 3\, M_P^2\, H(\arh)^2\,,    
\end{align}
when the inflaton and radiation energy densities become equal. To avoid affecting the BBN predictions, the reheating temperature $\Trh$ must satisfy $\Trh > T_\text{BBN} \simeq 4$~MeV~\cite{Sarkar:1995dd, Kawasaki:2000en, Hannestad:2004px, DeBernardis:2008zz, deSalas:2015glj, Hasegawa:2019jsa}.
%%%%%%%%%%%%%%%%%%%%%
\begin{figure}
\centering
\includegraphics[scale=0.37]{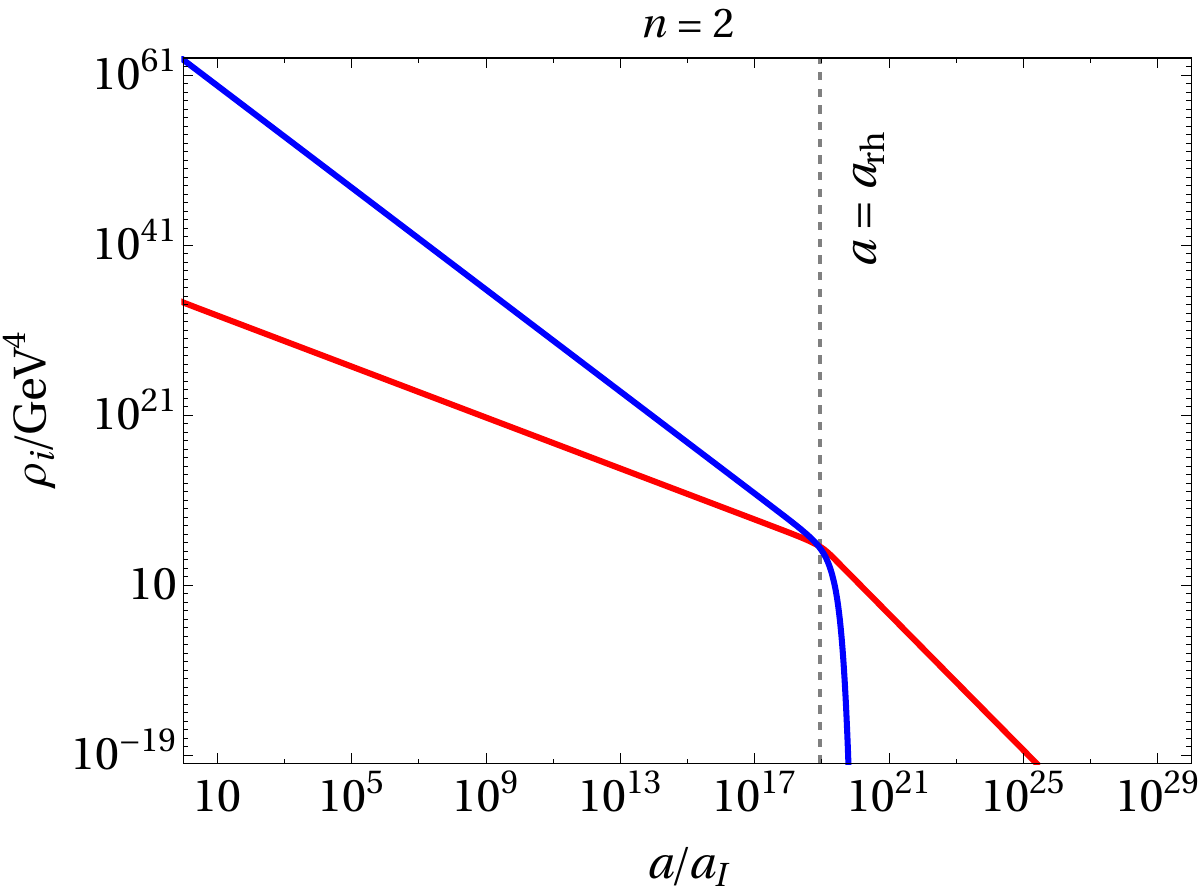}~\includegraphics[scale=0.37]{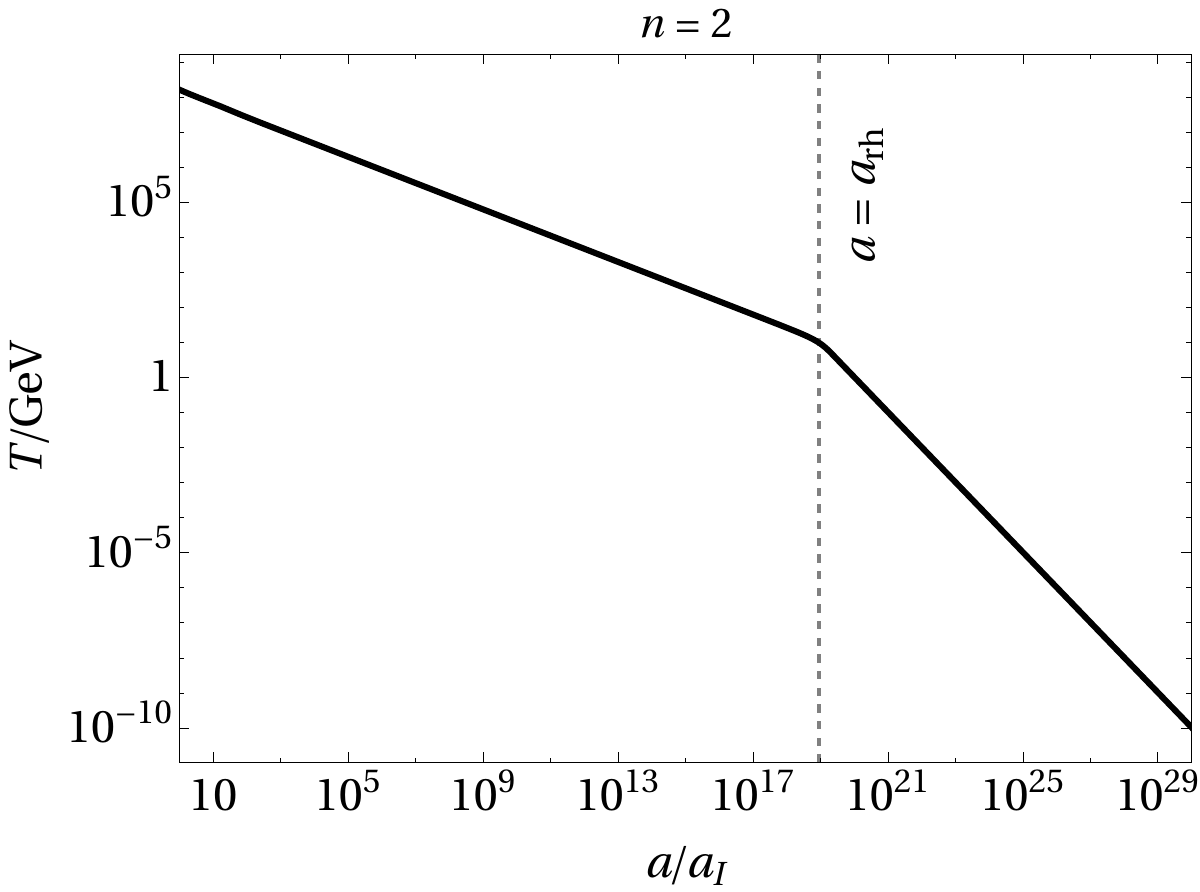}\\[10pt]
\includegraphics[scale=0.37]{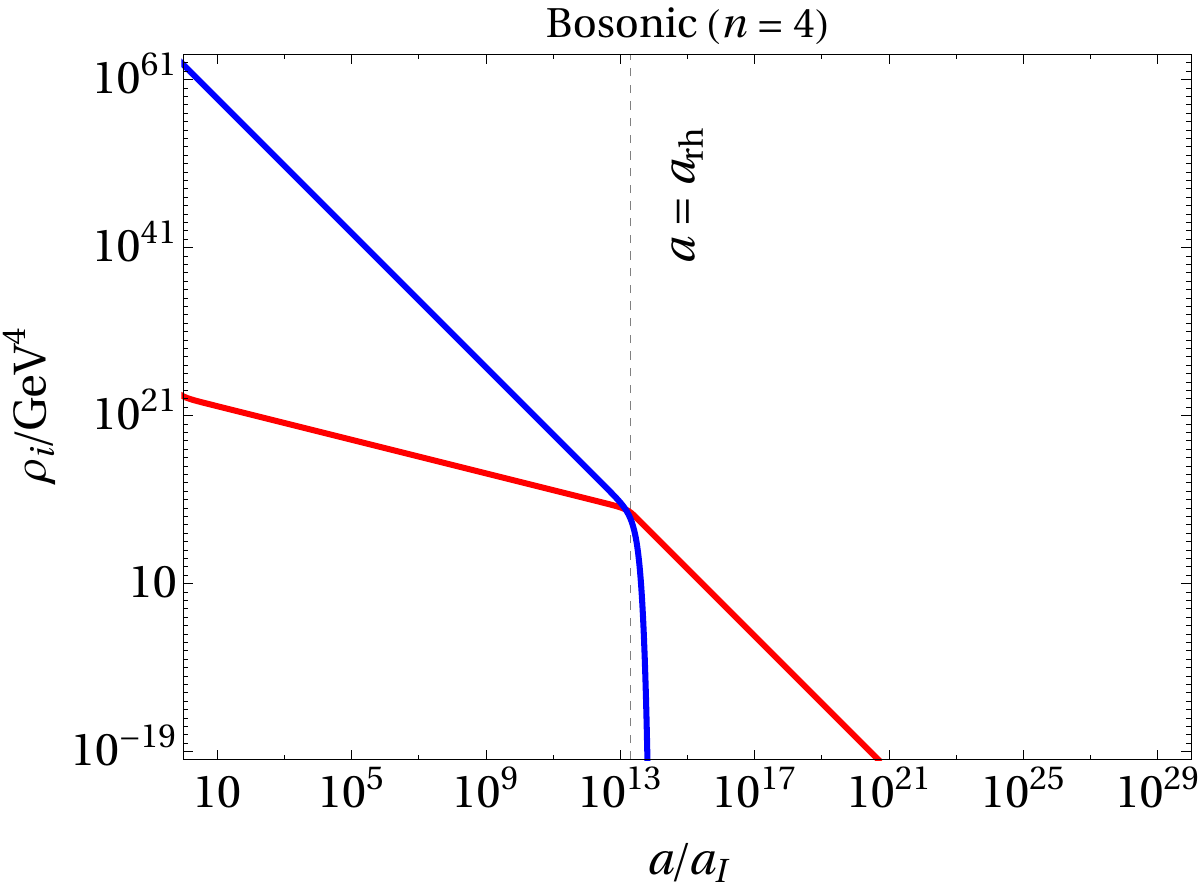}~\includegraphics[scale=0.37]{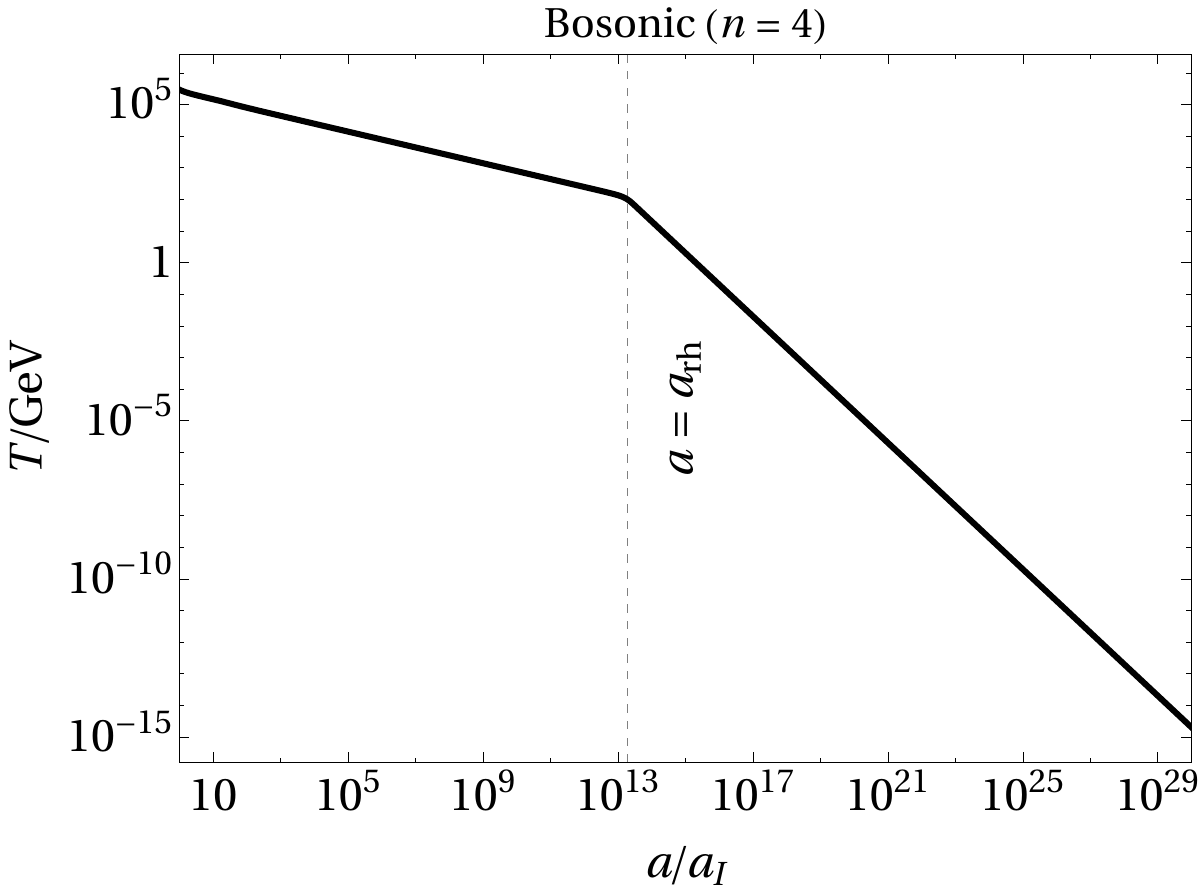}\\[10pt]
\includegraphics[scale=0.37]{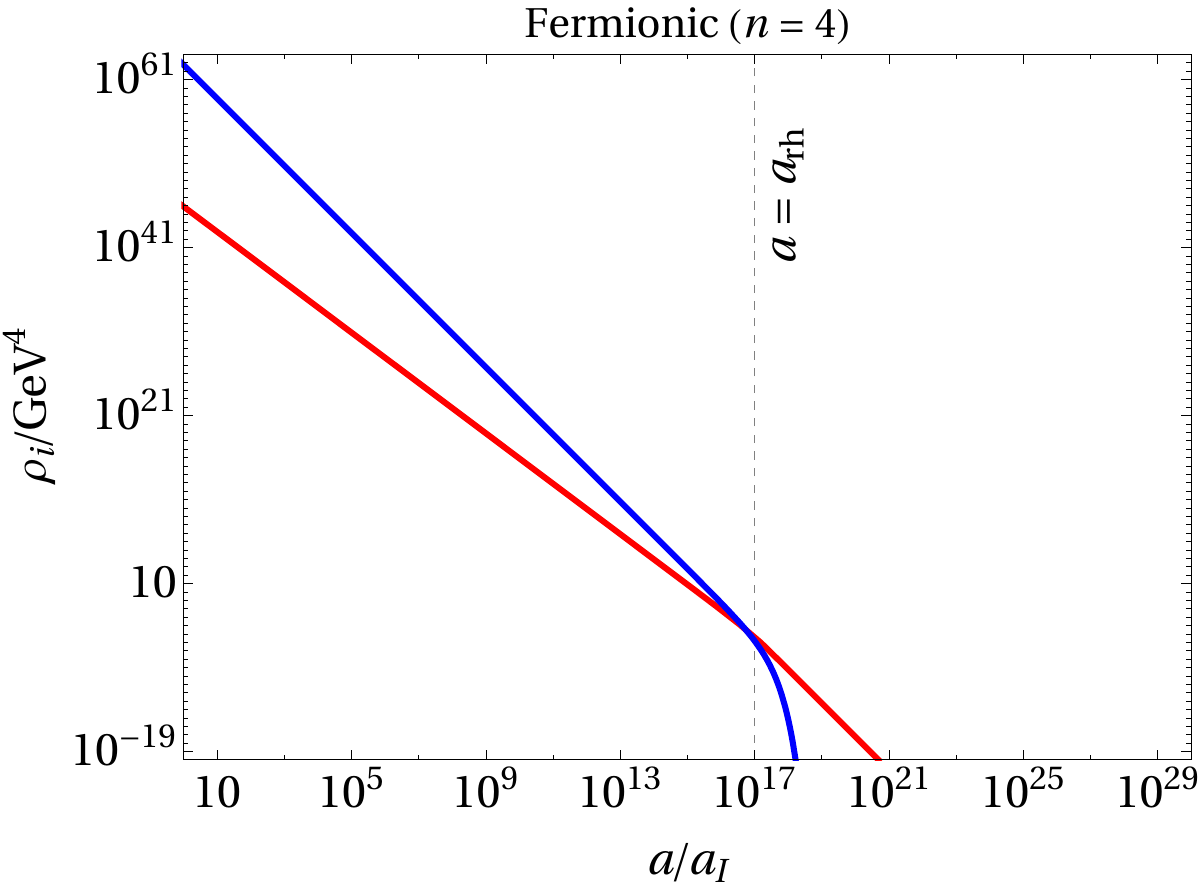}\includegraphics[scale=0.37]{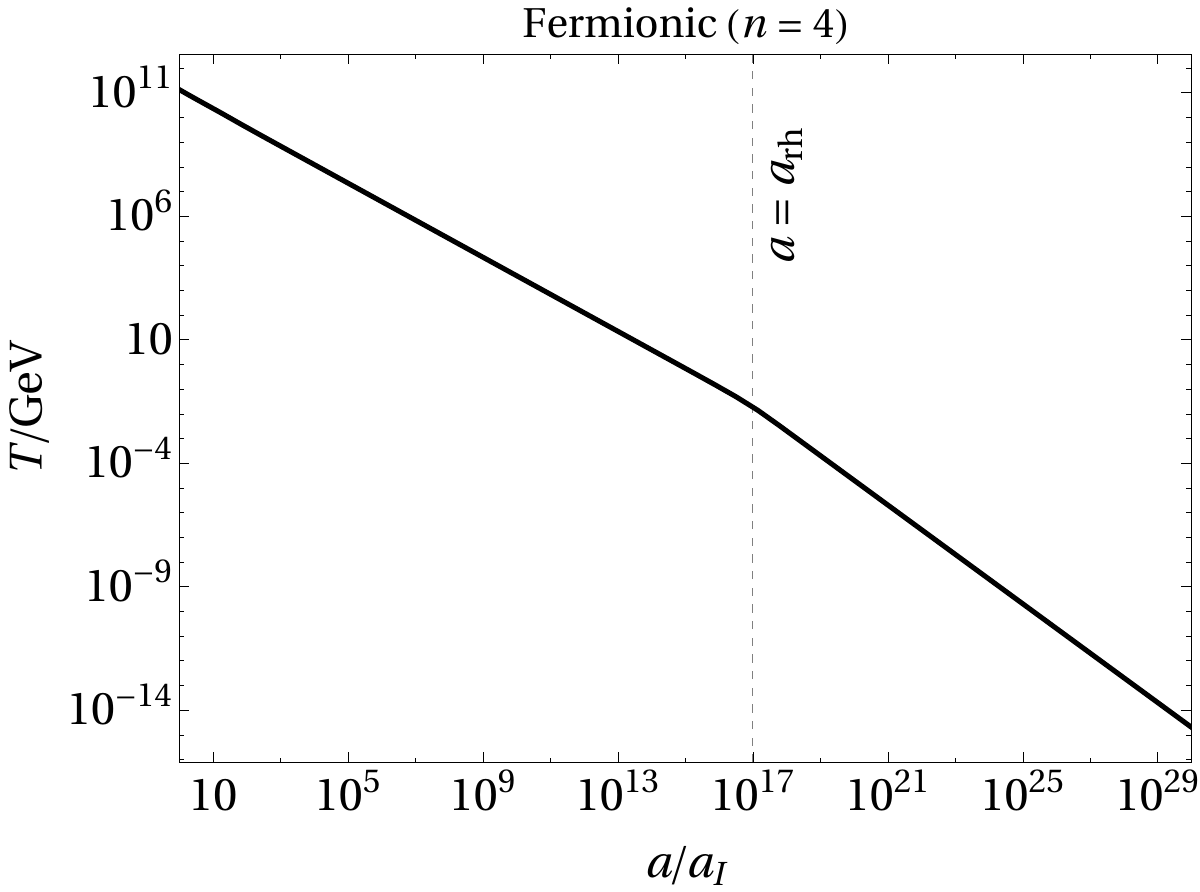}
\caption{Left column: Evolution of inflaton and radiation energy density as a function of scale factor for bosonic and fermionic reheating scenario, with different choices of $n$. Right column: Corresponding SM bath temperature as a function of scale factor. Here we considered $\lNP=1$ TeV, with $k=4$ and $\Trh=10$ GeV. The vertical dashed line in each figure shows the onset of radiation domination $(a=\arh)$.}
\label{fig:rho}
\end{figure}
%%%%%%%%%%%%%%%%%%%%%%%%%%%
\begin{figure}[htb!]
\centering
\includegraphics[scale=0.55]{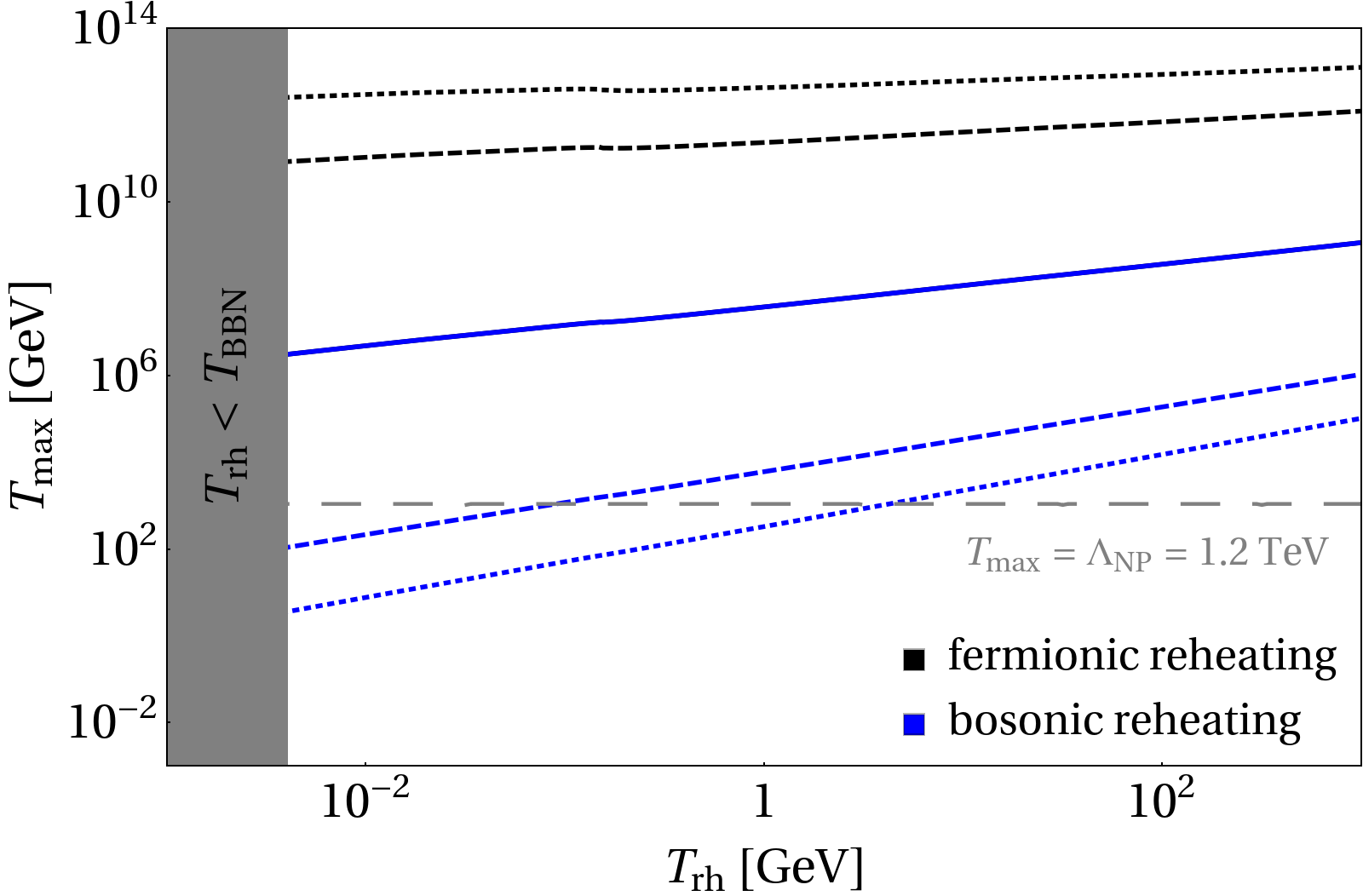}
\caption{Maximum temperature during reheating $\Tmax$, as a function of the reheating temperature $\Trh$. The solid, dashed and dotted curves correspond to $n=2,\,4,\,6$, respectively.}
\label{fig:tmaxtrh}
\end{figure}
%%%%%%%%%%%%%%%%%%%%%%
%%%%%%%%%%%%%%%%%%%%
\subsection{Reheating from perturbative inflaton decay}
\label{sec:inf-decay}
%%%%%%%%%%%%%%%%%%%
We consider reheating happens from the {\it perturbative} decay of the inflaton at the minimum of its potential. For two-body decay processes, we consider trilinear interactions between the inflaton $\phi$ and a pair of complex scalar doublets $\varphi$ (e.g., the Higgs boson doublet) or a pair of vector-like Dirac fermions $\Psi$, which are part of the SM bath\footnote{Gauge-invariant interaction between the inflaton and the SM fermions can be realized in a more realistic set-up. For example, the inflaton can be promoted to an electroweak multiplet~\cite{Chen:2010uc, Borah:2018rca} or electroweak gauge symmetry may not be restored at high temperatures~\cite{Meade:2018saz}.  We, however, remain agnostic about the microscopic model and write an effective inflaton-SM interaction as in Refs.~\cite{Kaneta:2019zgw,Garcia:2020wiy,Datta:2023pav}.}. The corresponding Lagrangian density reads
\begin{equation}\label{eq:int1}
\mathcal{L}_\text{int} \supset -\mu\, \phi\, |\varphi|^2 - y_\psi\, \overline{\Psi}\, \Psi\, \phi\,.    
\end{equation}
In writing the above interaction, we have assumed that the inflaton decays entirely into the visible sector fields that form the radiation bath. Moreover, we remain agnostic about the UV completion of the inflaton-SM coupling. In the following analysis, we will consider that each reheating scenario is present one at a time, and they can not occur simultaneously.
%%%%%%%%%%%%%%%%%%
\subsubsection{Fermionic channel}
\label{sec:fermionic}
%%%%%%%%%%%%%%%%%%
We first consider the scenario where the inflaton decays into a pair of fermions via the Yukawa interaction in Eq.~\eqref{eq:int1}, with a decay rate
\begin{equation} \label{eq:fer_gamma}
\Gp(a) = \frac{\yeff^2}{8\pi}\, m_\phi(a)\,,
\end{equation}
where the effective coupling $\yeff \ne y_\psi$ (for $n \neq 2$) is obtained after averaging over several oscillations~\cite{Shtanov:1994ce, Ichikawa:2008ne, Garcia:2020wiy}. The radiation energy density, following Eq.~\eqref{eq:rR}, evolves as~\cite{Bernal:2022wck,Barman:2023rpg,Xu:2024fjl}
\begin{equation} \label{eq:rR_fer}
\rR(a) \simeq \frac{3\, n}{7 - n}\, M_P^2\, \Gp(\arh)\, H(\arh) \left(\frac{\arh}{a}\right)^\frac{6 (n - 1)}{2 + n} \left[1 - \left(\frac{a_I}{a}\right)^\frac{2 (7 - n)}{2 + n}\right]\,.
\end{equation}
The corresponding temperature of the thermal bath reads
\begin{equation} \label{eq:Tevol}
T(a) \simeq \Trh \left(\frac{\arh}{a}\right)^\alpha,
\end{equation}
with
\begin{equation} \label{eq:Tfer}
\alpha =
\begin{cases}
\frac32\, \frac{n - 1}{n + 2} & \text{ for } n < 7\,,\\
1 & \text{ for } n > 7\,.
\end{cases}
\end{equation}
Here we assume that the SM bath thermaizes instantaneously\footnote{A detailed study of the thermalization of the inflaton decay products can be found, for example, in Refs.~\cite{Davidson:2000er, Kurkela:2011ti, Harigaya:2013vwa, Drees:2021lbm, Drees:2022vvn, Mukaida:2022bbo}.}. Trading the scale factor with temperature, the Hubble expansion rate during reheating (cf. Eq.~\eqref{eq:Hubble}) can be rewritten as
\begin{equation} \label{eq:Hevol}
H(T) \simeq H(\Trh) \left(\frac{T}{\Trh}\right)^{\frac{3\, n}{2 + n}\, \frac{1}{\alpha}}.
\end{equation}
%%%%%%%%%%%%%%%%%%%%
\subsubsection{Bosonic channel}
\label{sec:bosonic}
%%%%%%%%%%%%%%%%%%%%
If the inflaton decays solely into a pair of bosons through the trilinear scalar interaction in Eq.~\eqref{eq:int1}, the decay rate is given by
\begin{equation} \label{eq:bos_gamma}
\Gp(a) = \frac{\mueff^2}{8\pi\, m_\phi(a)}\,,
\end{equation}
where again the effective coupling $\mueff \ne \mu$ (if $n\neq2$) can be obtained after averaging over oscillations. Using a procedure similar to the previous case, the SM energy density scales as~\cite{Bernal:2022wck}
\begin{equation} \label{eq:rR_bos}
\rR(a) \simeq \frac{3\, n}{1 + 2\, n}\, M_P^2\, \Gp(\arh)\, H(\arh) \left(\frac{\arh}{a}\right)^\frac{6}{2 + n} \left[1 - \left(\frac{a_I}{a}\right)^\frac{2\, (1 + 2 n)}{2 + n}\right]\,,
\end{equation}
with which the SM temperature and the Hubble rate evolve similarly as Eqs.~\eqref{eq:Tevol} and~\eqref{eq:Hevol}, respectively, with
\begin{equation} \label{eq:TBos}
\alpha = \frac32\, \frac{1}{n + 2}\,,
\end{equation}
during reheating. For $n=2$ the inflaton equation of state is that of a pressureless non-relativistic matter $(w=0)$ and from Eq.~\eqref{eq:Tevol} we see that we reproduce the familiar scale factor-temperature dependence for inflaton oscillating in a quadratic potential. Finally, the maximum temperature of the bath during reheating reads
\begin{align}\label{eq:Tmax1}
\Tmax\simeq \Trh
\begin{cases}
\left(\arh/a_I\right)^\frac{3\,(n-1)}{2\,(n+2)} & \text{fermionic reheating}\,,
\\[10pt]
\left(\arh/a_I\right)^\frac{3}{2\,(n+2)} & \text{bosonic reheating}\,,
\end{cases}
\end{align}
which can be several orders of magnitude larger than $\Trh$, for $\arh\gg a_I$. 

In the left column of Fig.~\ref{fig:rho}, we show the evolution of inflaton (blue) and radiation (red) energy densities as a function of the scale factor for both bosonic and fermionic reheating cases. Here we have fixed $\lNP=1$ TeV, $\Trh=10$ GeV and $k=4$, such that corresponding DM-SM operators are of dimension $d=7$. Corresponding temperature evolution of the SM radiation bath is shown in the right column, which scales following Eq.~\eqref{eq:Tevol}. For example, with $n=2$, in either cases, $T\propto a^{-3/8}$, while for $n=4$, $T\propto a^{-3/4}$ for fermionic and $T\propto a^{-1/4}$ for bosonic case, respectively. Note that, the inflaton decay width for fermionic reheating $\propto m_\phi(a)$, while for bosonic final states it is $\propto 1/m_\phi(a)$. As a consequence, the reheating process becomes more efficient over time for the bosonic final states than for the fermionic final states for $n>2$, since the inflaton mass $m_\phi(a)$ is a decreasing function of time. Now, the effective description in Eq.~\eqref{eq:gamma-DM} is valid for the duration of reheating provided, $\lNP\gtrsim\Tmax$ \cite{Garcia:2020eof}. On the other hand, since our goal is to explore reheating scenarios at the collider with $\sqrt{s}=1$ TeV, therefore we prefer to have $\lNP\gtrsim 1$ TeV\footnote{At colliders, the EFT framework breaks down for $\sqrt{s} > \lNP$.}. To be more precise, for the collider analysis, we will fix $\lNP=1.2$ TeV. Consequently, it is important to verify if the hierarchy $\lNP\gtrsim\Tmax$ is maintained for both bosonic and fermionic reheating scenarios. In Fig.~\ref{fig:tmaxtrh}, we show the maximum temperature $\Tmax$ during reheating as a function of $\Trh$ for different choices of $n$, considering both bosonic and fermionic cases. We see, $\lNP\gtrsim\Tmax$ is satisfied for bosonic reheating only, and for $n=4,\,6$ with $\Trh$ around the MeV scale, satisfying the BBN bound. For fermionic reheating, $\Tmax$ is always much higher than $\lNP$, while for larger $n$ in case of bosnic reheating, $\Tmax$ decreases with $\Trh$. This can be simply understood from Eq.~\eqref{eq:Tmax1}, where we see, $\Tmax$ is approximately constant for fermionic reheating and for bosonic, $\Tmax\propto 1/\Trh$, for $n\gg 1$. Therefore, in order to maintain $\lNP\gtrsim\Tmax$, together with $\lNP=1.2$ TeV, we will stick to the bosonic reheating framework. 

Before closing this section it is worth mentioning that in this work we focus on low-reheating-temperature scenarios (as that will provide large signal significance at the collider), typically around MeV scale and assume small inflaton couplings to daughter particles, which makes preheating effects inefficient. Therefore, throughout this work, we will concentrate on the perturbative reheating scenario. In any case, to fully deplete the inflaton energy, perturbative decay through trilinear couplings between the inflaton and daughter particles is still necessary, which is expected to occur in the final stage of the heating process after inflation, when most of the energy is transferred from the inflaton to the bath  particles~\cite{Lozanov:2016hid,Lozanov:2017hjm,Maity:2018qhi,Saha:2020bis,Antusch:2020iyq}. For $n \gtrsim 8$ (or equivalently $w\gtrsim 0.65$), gravitational reheating becomes efficient~\cite{Haque:2022kez, Clery:2022wib, Co:2022bgh, Haque:2023yra}, and in certain cases may surpass perturbative reheating, depending on the inflaton-matter coupling, as demonstrated in Refs.~\cite{Haque:2022kez, Haque:2023yra}. Typically, when $w\gtrsim 0.65$, gravitational reheating alone can be sufficient to reheat the Universe without requiring any input from perturbative processes. However, since our focus is on perturbative reheating, we will restrict our analysis to $n<8$. 
%%%%%%%%%%%%%%%%%%%
\subsection{Dark matter genesis during reheating}
\label{sec:dm}
%%%%%%%%%%%%%%%%%%%
The reheating via fermionic or bosonic channel results in the production of the SM radiation bath. We consider production of DM from the SM bath during the reheating via a UV freeze-in. The DM number can then be tracked by solving the Boltzmann equation for its number density $\ndm$
\begin{equation} \label{eq:BEDM0}
\frac{d\ndm}{dt} + 3\, H\, \ndm = \gamma\,,
\end{equation}
where $\gamma$ is the DM production reaction rate density, which we parametrize as~\cite{Elahi:2014fsa,Bernal:2019mhf,Kaneta:2019zgw,Barman:2023ktz}
\begin{align}\label{eq:gamma-DM}
& \gamma=\frac{T^{k+6}}{\Lambda_{\rm NP}^{k+2}}\,.    
\end{align}
Here $k=2\,(d-5)$, where $d$ is the dimension of the relevant effective DM-SM operator ($d \geq 5$), and $\lNP$ is identified with the beyond the SM NP scale of the microscopic model under consideration. The scale of inflation $\Lambda$ is seemingly uncorrelated to $\lNP$. We adopt this parametrization for the DM reaction density to obtain an approximate analytical expression for DM yield. Numerically, while solving the Boltzmann equation, we compute the reaction density using the expressions for annihilation cross-sections provided in Appendix.~\ref{sec:cs}. As the SM entropy is not conserved during reheating due to the annihilation of the inflaton in SM particles, it is convenient to introduce a comoving number density $\Ndm \equiv \ndm\, a^3$, and therefore Eq.~\eqref{eq:BEDM0} can be rewritten as
\begin{equation} \label{eq:BEDM}
\frac{d\Ndm}{da} = \frac{a^2\,\gamma}{H}\,,
\end{equation}
which has to be numerically solved together with Eqs.~\eqref{eq:drhodt} and~\eqref{eq:rR_fer} or \eqref{eq:rR_bos}, taking the initial condition $\Ndm(a_I) = 0$. To fit the whole observed DM relic density, it is required that
\begin{equation} \label{eq:obsyield}
Y_0\, \mdm = \Omega h^2 \, \frac{1}{s_0}\,\frac{\rho_c}{h^2} \simeq 4.3 \times 10^{-10}~\text{GeV},
\end{equation}
where $Y_0 \equiv Y(T_0)$ and $Y(T) \equiv \ndm(T)/s(T)$, with $s$ being the SM entropy density defined by
\begin{equation}
s(T) = \frac{2\pi^2}{45}\, \gss(T)\, T^3,
\end{equation}
and $\gss(T)$ being the number of relativistic degrees of freedom contributing to the SM entropy. Furthermore, $\rho_c \simeq 1.05 \times 10^{-5}\, h^2$~GeV/cm$^3$ is the critical energy density, $s_0\simeq 2.69 \times 10^3$~cm$^{-3}$ the present entropy density~\cite{ParticleDataGroup:2022pth}, and $\Omega h^2 \simeq 0.12$ the observed abundance of DM relics~\cite{Planck:2018vyg}. Now, depending on the DM mass-scale, two situations may appear. In case $\mdm\ll\Trh$, the DM is produced at the end of reheating, with a DM number
\begin{align}\label{eq:Ndm1}
& \Ndm(\arh)\simeq\frac{2\,\sqrt{10}\,(n+2)}{\pi\,\gss(\Trh)}\,\frac{M_P\,\Trh^{k+4}}{\lNP^{k+2}}\,\arh^3
\nonumber\\&\times
\begin{cases}
\frac{1}{k-n\,(k+2)+10}\,\left[1-\left(a_I/\arh\right)^{\frac{3\,(k+10-n\,(k+2)}{2n+4}}\right]\,, & \text{fermionic reheating}
\\[10pt]
\frac{1}{4\,n-k-2}\,\left[1-\left(a_I/\arh\right)^\frac{12n-3k-6}{2n+4}\right]\,, & \text{bosonic reheating}\,,
\end{cases}
\end{align}
and corresponding yield
\begin{align}\label{eq:ydm1}
& Y(\arh)\simeq\frac{45\,\sqrt{10}\,(n+2)}{\pi^3\,\gss(\Trh)^{3/2}}\,\frac{M_P\,\Trh^{k+1}}{\lNP^{k+2}}
\nonumber\\&\times
\begin{cases}
\frac{1}{k-n\,(k+2)+10}\,\left[1-\left(a_I/\arh\right)^{\frac{3\,(k+10-n\,(k+2)}{2n+4}}\right]\,, & \text{fermionic reheating}
\\[10pt]
\frac{1}{4\,n-k-2}\,\left[1-\left(a_I/\arh\right)^\frac{12n-3k-6}{2n+4}\right]\,, & \text{bosonic reheating}\,.
\end{cases}
\end{align}
The ratio of the scale factors $a_I/\arh$ can be replaced using Eq.~\eqref{eq:rpsol} as
\begin{align}
& a_I=\arh\times\left(\frac{H(\arh)}{H(a_I)}\right)^\frac{n+2}{3n}\,.    
\end{align}

On the other hand, for $\Trh<\mdm\ll\Tmax$, DM is produced {\it during} reheating. In this case, the DM number density can be estimated by integrating Eq.~\eqref{eq:BEDM} from the $a_I$ to $\adm=a(T=\mdm)$, where
\begin{align}\label{eq:admarh}
& \adm = \arh\times\left(\frac{\Trh}{\mdm}\right)^{1/\alpha}\,,  
\end{align}
as can be obtained using Eq.~\eqref{eq:Tevol}. In this case, DM number at $a=\adm$ is given by
\begin{align}\label{eq:Ndm2}
& \Ndm(\adm)\simeq\frac{2\,\sqrt{10}\,(n+2)}{\pi\,\gss(\mdm)}\,\frac{M_P\,\Trh^{k+4}}{\lNP^{k+2}}\,\arh^3
\nonumber\\&\times
\begin{cases}
\left(\Trh/\mdm\right)^\frac{k-n(k+2)+10}{n-1}\,\frac{1}{k-n\,(k+2)+10}\,\left[1-\left(a_I/\adm\right)^{\frac{3\,(k+10-n\,(k+2)}{2n+4}}\right]\,, & \text{fermionic reheating}
\\[10pt]
\left(\Trh/\mdm\right)^{4n-k-2}\,\frac{1}{4\,n-k-2}\,\left[1-\left(a_I/\adm\right)^\frac{12n-3k-6}{2n+4}\right]\,, & \text{bosonic reheating}\,,
\end{cases}
\end{align}
with corresponding yield
\begin{align}\label{eq:ydm2}
& Y(\adm)\simeq\frac{45\,\sqrt{10}\,(n+2)}{\pi^3\,\gss^{3/2}(\mdm)}\,\frac{M_P\,\Trh^{k+4}}{\mdm^3\,\lNP^{k+2}}
\nonumber\\&\times
\begin{cases}
\left(\Trh/\mdm\right)^\frac{6+k-n\,(k+4)}{n-1}\,\frac{1}{k-n\,(k+2)+10}\,\left[1-\left(a_I/\adm\right)^{\frac{3\,(k+10-n\,(k+2)}{2n+4}}\right]\,, & \text{fermionic reheating}
\\[10pt]
\left(\Trh/\mdm\right)^{2n-k-6}\,\frac{1}{4\,n-k-2}\,\left[1-\left(a_I/\adm\right)^\frac{12n-3k-6}{2n+4}\right]\,, & \text{bosonic reheating}\,.
\end{cases}
\end{align}
The final yield at the end of reheating, in this case, reads
\begin{align}
& Y(\arh) = Y(\adm)\, \left(\frac{\adm}{\arh} \right)^3\frac{s(\adm)}{s(\arh)} \simeq Y(\adm)\,\left(\frac{\mdm}{\Trh}\right)^3\,\left(\frac{\adm}{\arh}\right)^3\,,
\end{align}
up to some order of the ratio of $\gss$. Here $\adm/\arh$ is given by Eq.~\eqref{eq:admarh} and $Y(\adm)=\Ndm(\adm)/\left(\adm^3\,s(\mdm)\right)$, in order to include the effect of entropy injection during reheating.
%%%%%%%%%%%%%%%%%%%%%%%%%%%%
\begin{figure}[htb!]
\centering
\includegraphics[scale=0.2]{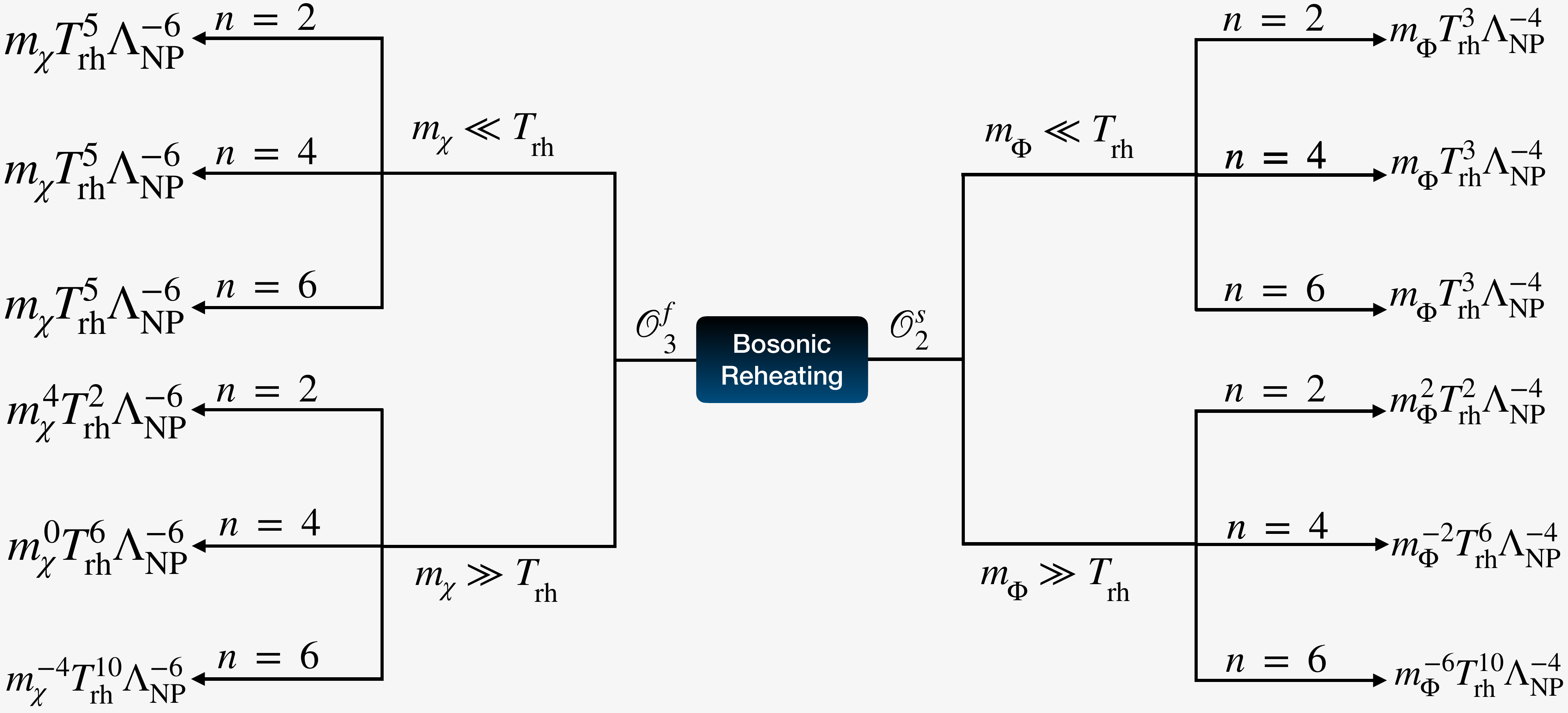}
\caption{Summary of parameter dependence of relic abundance for scalar $(\mathcal{O}_2^s)$ and fermionc $(\mathcal{O}_3^f)$ dark matter, considering the bosonic reheating scenario for $n=2,\,4,\,6$.}
\label{fig:yields}
\end{figure}
%%%%%%%%%%%%%%%%%%%%%%%%%%%%
\begin{figure}[htb!]
\centering
\includegraphics[scale=0.25]{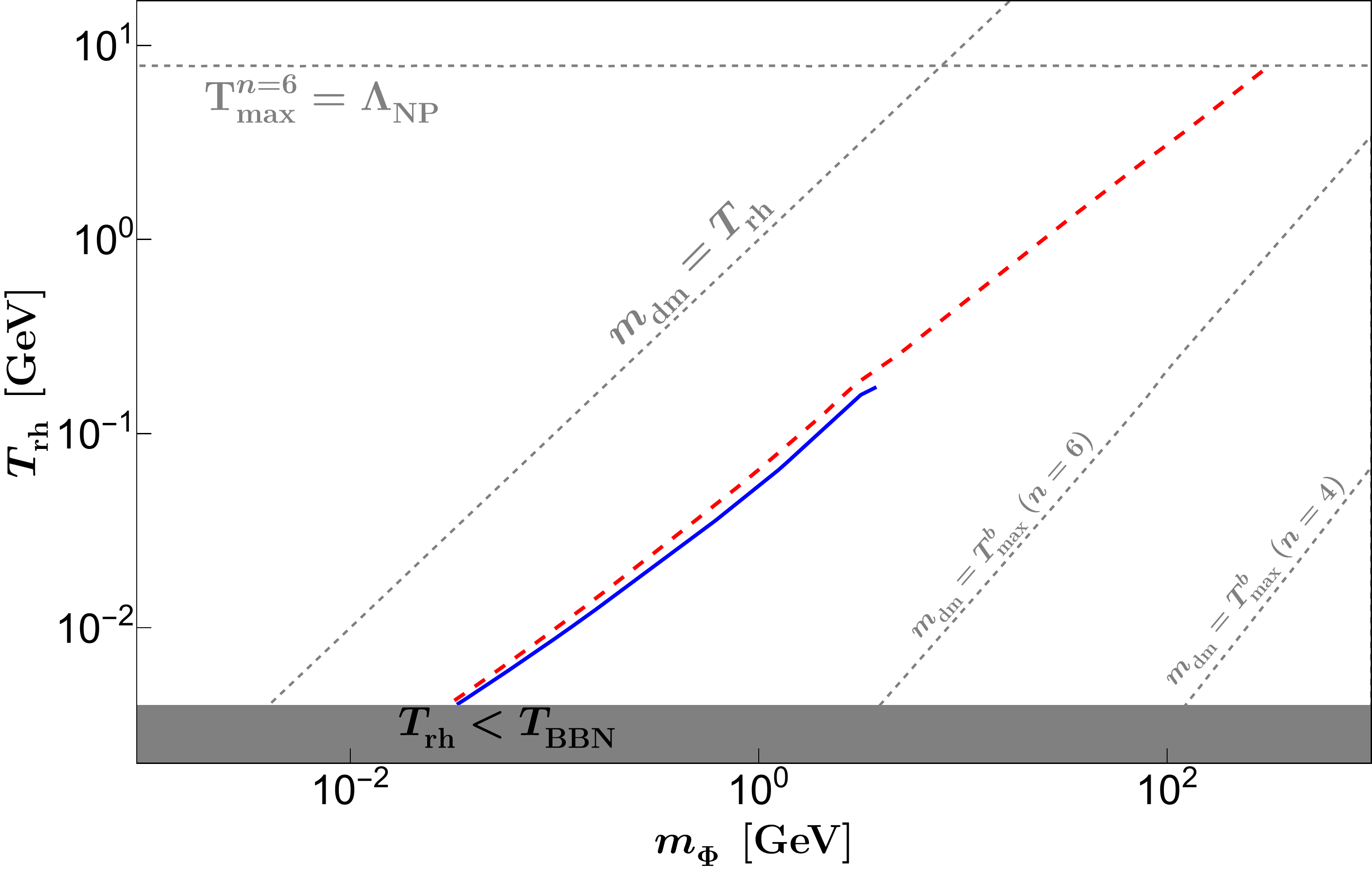}\\[10pt]
\includegraphics[scale=0.25]{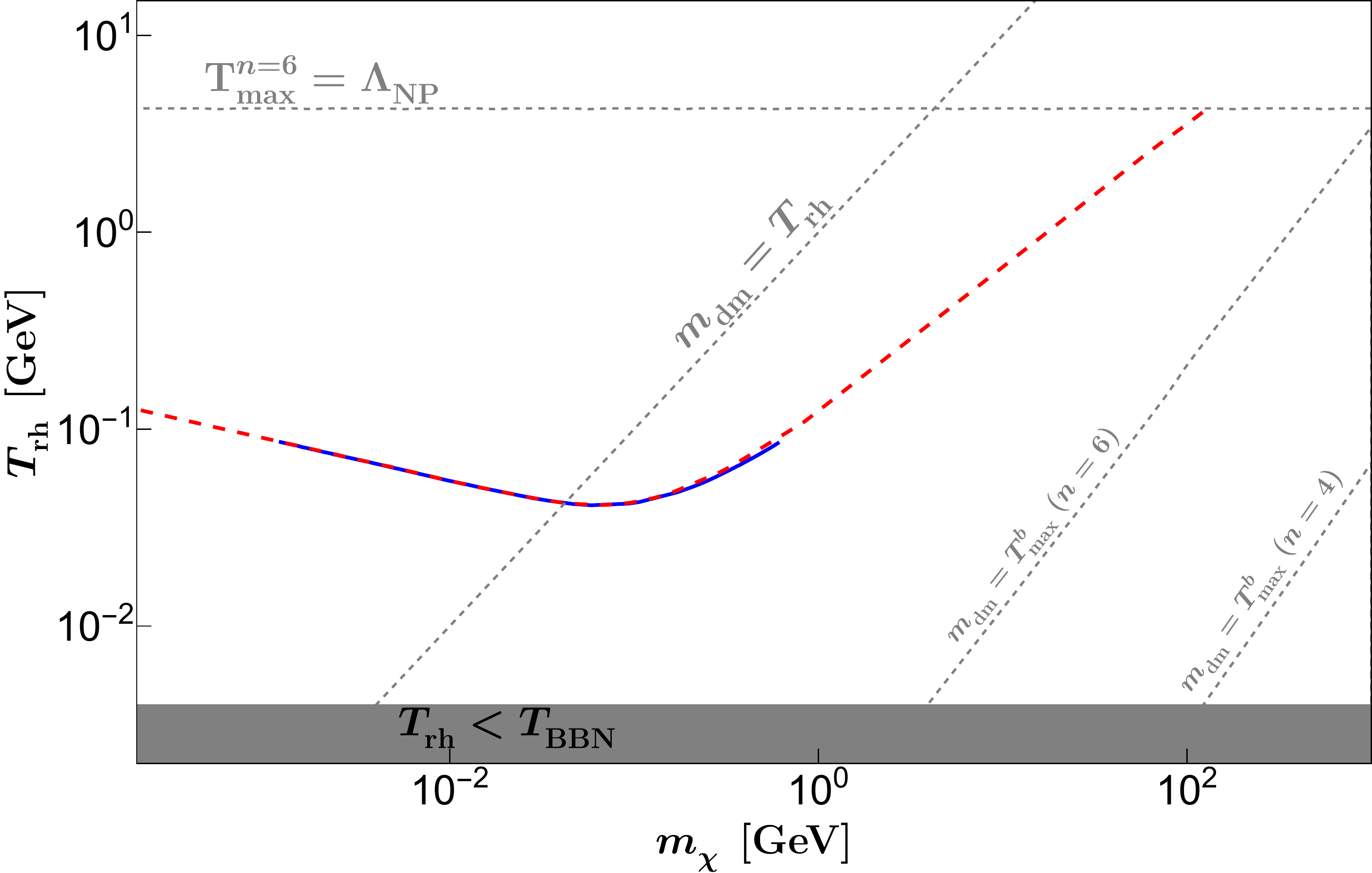}
\caption{Viable parameter space for scalar (top) and fermionic DM (bottom), satisfying the observed relic abundance for the bosonic reheating scenario. The blue solid and red dashed lines correspond to $n=4$ and $n=6$, respectively. The gray shaded region is forbidden from BBN bound on $\Trh$. \textcolor{black}{We have fixed $\Lambda_{\mathrm{NP}} = 2.0~(1.2)$ TeV for the scalar (fermionic) DM scenario, ensuring consistency with indirect search constraints and maintaining the validity of the EFT approach in the collider context, where $\sqrt{s} < \Lambda_{\mathrm{NP}}$.}}
\label{fig:mdm-trh}
\end{figure}
%%%%%%%%%%%%%%%%%%%%%%%%%%%%

The dependence of final DM yield on $\Trh,\,\mdm$ and $\lNP$ can be analytically obtained using Eq.~\eqref{eq:ydm1} and \eqref{eq:ydm2}. For better understanding, we have gathered them in Fig.~\ref{fig:yields}, for both types of reheating and for $n=2,\,4,\,6$. We will be concentrating on the bosonic reheating part and consider $n=\{4,\,6\}$ as explained in the last paragraph. Now, for $\mdm\ll\Trh$, from Eq.~\eqref{eq:ydm1}, we find,
\begin{align}
& \frac{\Trh}{\mdm}\propto\left(\frac{\lNP}{\mdm}\right)^\frac{k+2}{k+1}\,, 
\end{align}
where we have used the fact that $a_I/\arh\ll 1$. This is clearly visible from the bottom panel of Fig.~\ref{fig:mdm-trh} for fermionic DM, where we note the negative slope for each contour satisfying observed relic abundance. For scalar DM, on the other hand, right relic abundance in this limit is produced for $\Trh<T_{\rm BBN}$ and hence forbidden from the BBN bound. For $\Trh\ll\mdm\ll\Tmax$, it is difficult to find an exact analytical expression from Eq.~\eqref{eq:ydm2}. One can however approximately obtain,
\begin{align}
& \frac{\Trh}{\mdm}\propto\left(\frac{\mdm^p}{\lNP^{k+2}}\right)^\frac{1}{2n-k-2}\,,   
\end{align}
with $p=\left(k+2-n(k+6)\right)/n$, and we have considered $(k+2)/2n\ll 1$, along with $n>1$. This shows, corresponding slope for the allowed parameter space in $\Trh-\mdm$ is positive, which is again confirmed from both the plots in Fig.~\ref{fig:mdm-trh}. Therefore, for scalar DM, the relic density allowed parameter space is realizable only for $\Trh\ll\mdm\ll\Tmax$, whereas for fermionic DM, correct abundance can be produced for both $\mdm\ll\Trh$ and $\Trh\ll\mdm\ll\Tmax$. The maximum reheating temperature, for each $n$, that provides right DM abundance also corresponds to maximum temperature of the thermal bath $\Tmax=\lNP$, above which the effective DM-SM description remains valid no more. In the figure we only show the $\Tmax=\lNP$ contour for $n=6$, the same also exists for $n=4$ but at a much lower $\Trh$. Note that, for instantaneous reheating (sudden decay approximation), it is not possible to produce DM with mass $\mdm>\Trh$, as the maximum available temperature to the radiation bath is $\Trh$ itself. This is what we realized in~\cite{Barman:2024nhr}, where the scalar DM, produced during instantaneous reheating, is completely ruled out from the BBN bound. The inclusion of non-instantaneous inflaton decay therefore helps in resurrecting the relic density allowed parameter space for the scalar DM on one hand, while for fermionic DM it opens up more parameter space, allowing heavier DM mass, depending on the choice of $n$ and $\Trh$. Before moving on we would like to mention that the relic density allowed parameter space for vector DM turns out to be similar to that of the scalar DM. For the same mass, vector DM corresponding to the operator $\mathcal{O}^{V}_{2}$ requires about $\mathcal{O}(1)$ lower $\Trh$ compared to that of scalar to produce the observed abundance, since in the former case, there is larger number of degrees of freedom. We, therefore, refrain from showing the resulting parameter space explicitly. For the operator $\mathcal{O}^{VV}_{4}$, collider probe of vector DM is limited considering all the phenomenological constraints.

%%%%%%%%%%%%%%%%%%%%%%%%%%%%%%%%%%%%%%%%%%%%%%%%%%%%%%%%%%%%%%%%%%%
\subsection{Limits from indirect DM search}
%%%%%%%%%%%%%%%%%%%%%%%%%%%%%%%%%%%%%%%%%%%%%%%%%%%%%%%%%%%%%%%%%%%%
% In this work, we consider an effective interaction between photons and DMs. As the DM phenomenology is based on the UV-FI mechanism, which typically requires a feeble interaction cross-section to achieve the correct relic abundance-often beyond the sensitivity of current experiments. However, in this scenario, the cross-section is not as suppressed as in the standard UV-FI case, implying a potential impact on the parameter space that may be probed by DM detection experiments. 
\textcolor{black}{
The relic density allowed parameter space in the present framework is tightly constrained from Fermi-LAT~\cite{Fermi-LAT:2015kyq}, H.E.S.S~\cite{HESS:2020zwn}, and Planck~\cite{Slatyer:2015jla} observations that put an upper bound on the DM annihilation cross-section, depending on the DM mass. Here, the thermally averaged DM-SM cross section can be sizable because we have $\lNP\sim 1$ TeV and $\Trh\sim\mathcal{O}$(MeV), to satisfy the DM abundance via freeze-in. These observations probe different ranges of DM mass through $\gamma \gamma$ annihilation. The Fermi-LAT experiment constrains the thermally averaged cross-section in the mass range of 200 MeV to 400 GeV, while HESS sets limits for higher masses, spanning from 200 GeV to 40 TeV. On the other hand, Planck provides constraints over a broader range\---covering DM masses from 10 MeV to 10 TeV.
We have translated bounds on DM annihilation cross-section into bounds on $\lNP$ and $\mdm$. }

%%%%%%%%%%%%%%%%%%%%%%%%%%%%%%
\begin{figure}[htb!]
\centering
\includegraphics[width=0.45\linewidth]{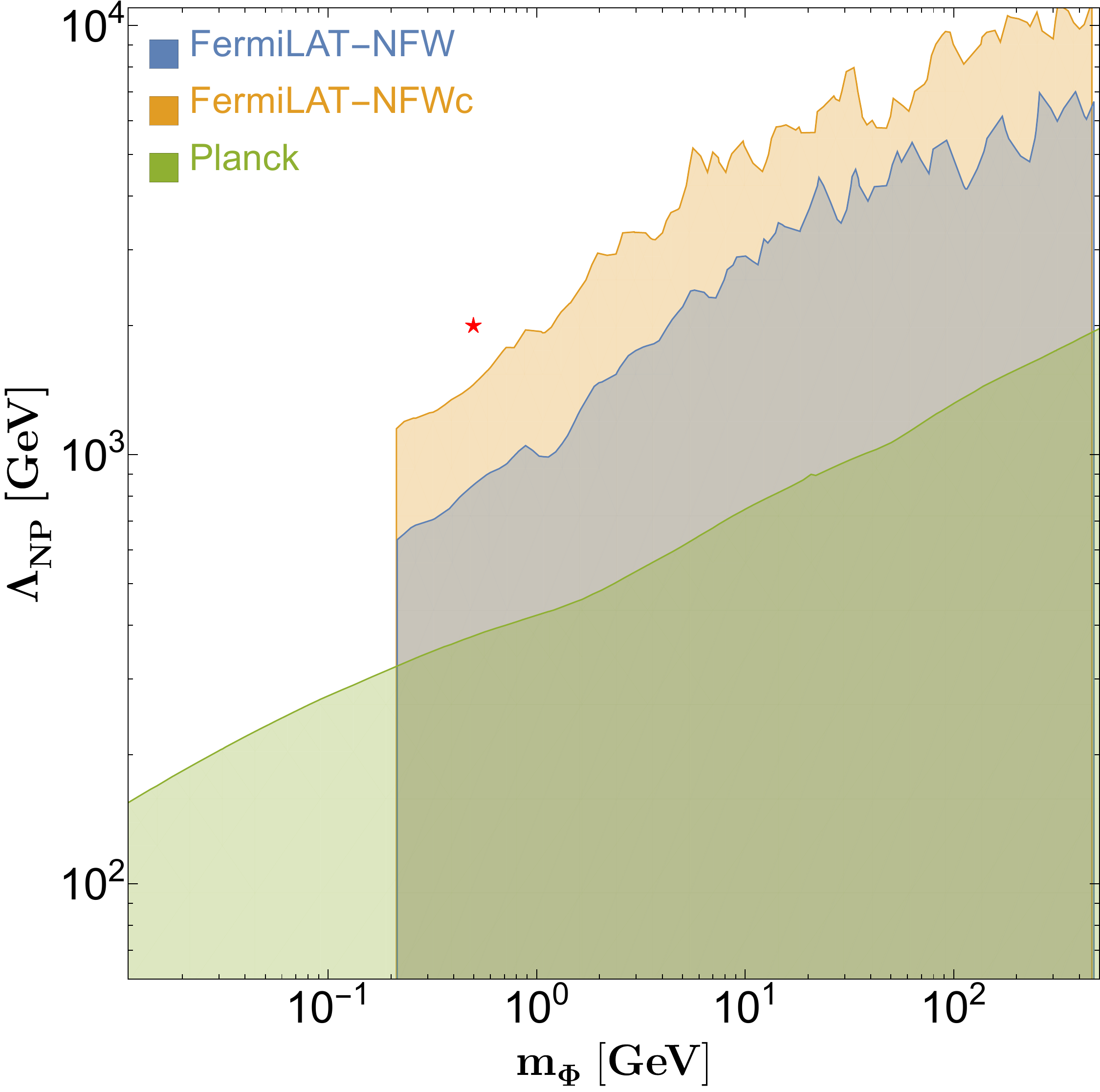}\quad
\includegraphics[width=0.45\linewidth]{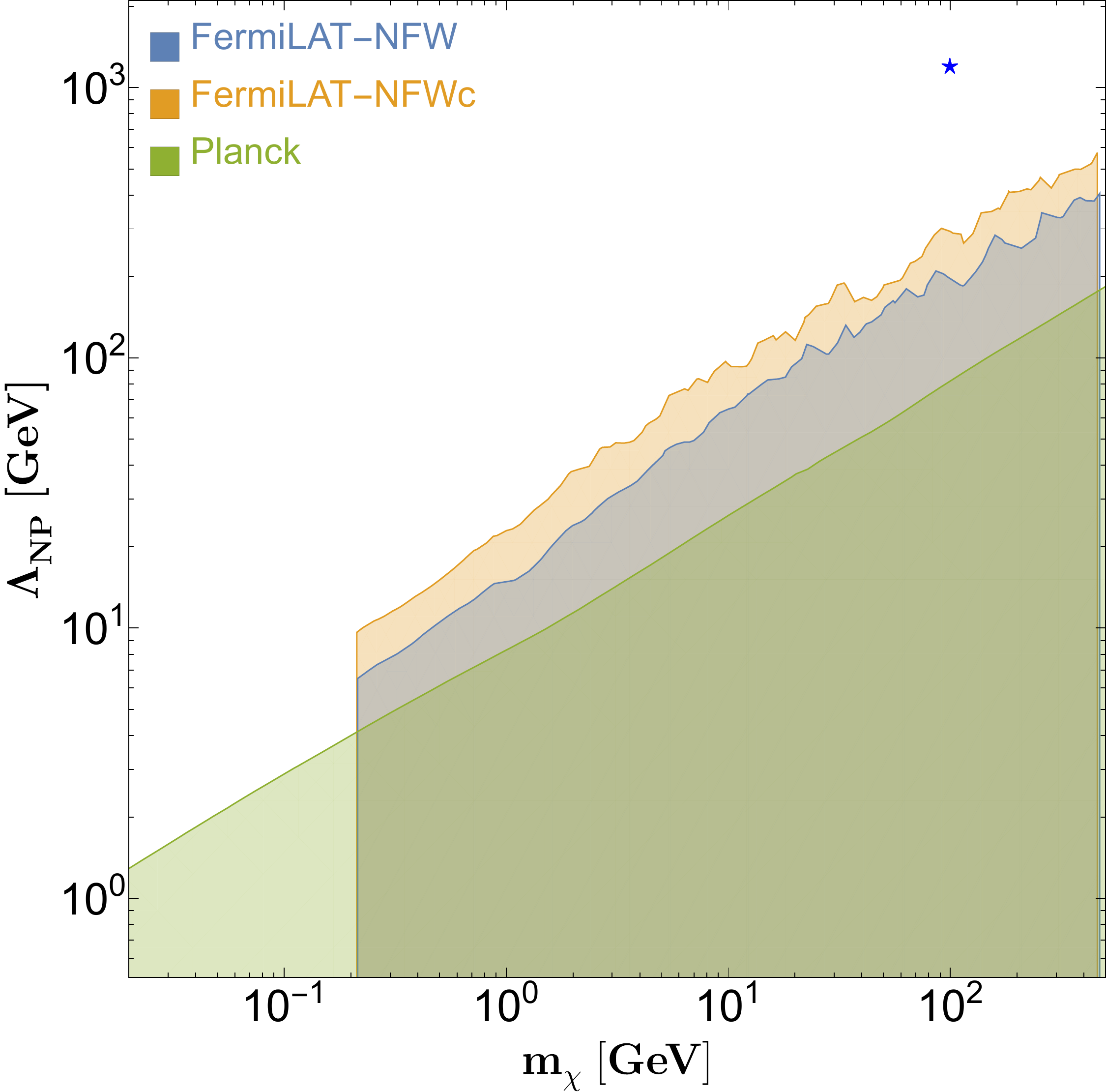}
\caption{The figures depict the indirect search constraints on the thermally averaged annihilation cross-section, $\langle\sigma v\rangle_{\rm DM~DM\to\gamma\gamma}$, based on current observational data, presented in the $\mdm-\lNP$ parameter plane for scalar (left) and fermionic (right) DM. The red (blue) star denotes the benchmark point utilized in the collider analysis for scalar (fermionic) DM.}
\label{fig:ID}
\end{figure}
%%%%%%%%%%%%%%%%

\textcolor{black}{In Fig.~\ref{fig:ID}, we show the constraints on scalar DM (left panel) and fermionic DM (right panel), annihilating into a photon pair. All the shaded regions are excluded. For DM mass below 200 MeV, Planck provides the relevant constraint in $\lNP-m_{\rm DM}$ plane. In the case of scalar DM with NP scale of $\lNP = 2~\rm TeV$, masses $m_{\Phi} > 1~\rm GeV$ are ruled out, whereas the constraint on $\lNP$ for fermionic DM is relaxed by two orders of magnitude. This is because scalar DM interacts via dimension-six operator, while fermionic DM involves dimension-seven operator, which is more suppressed due to their higher dimensionality. Therefore, the corresponding parameter space for fermionic DM remains viable. Based on these constraints, we select benchmark points from the allowed parameter region consistent with both relic abundance and indirect detection limits for the collider analysis in the following section.}

%%%%%%%%%%%%%%%%%%%%%%
\section{Testing reheating dynamics at lepton collider}
\label{sec:collider}
%%%%%%%%%%%%%%%%%%%%%%
We turn to discuss the prospects of detecting  DM produced during the epoch of reheating through the effective interactions in Eq.~\eqref{eq:eft-ops}. The relevant Feynman diagrams leading to such DM production at the electron-positron colliders are shown in Fig.~\ref{fig:ldm-cd}. As we know, the DM remains undetectable on the collider front. The conservation of four-momentum leads this discrepancy to appear what is known as the ``missing energy''. The detection of any excess in such missing energy events over the SM backgrounds, therefore, indicates the possibility of discovering DM at the colliders. Now, due to the absence of the QCD background, interactions at lepton colliders occur in a much cleaner environment compared to hadron colliders. Additionally, the availability of beam polarization is advantageous, as it can significantly reduce the SM background and/or enhance the NP signal. We thus perform signal-background analysis considering the proposed International Linear Collider (ILC)~\cite{Bambade:2019fyw}, for a center-of-mass (CM) energy of $\sqrt s=1$ TeV.
%%%%%%%%%%%%%%%%%%%%%%%%%%%%%%%%
\subsection{Mono-$\gamma$ signal}
\label{sec:ma}
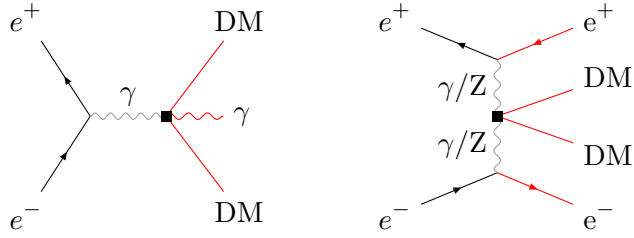
\begin{figure}[htb!]
\centering
\begin{tikzpicture}%[baseline={(current bounding box.center)},style={scale=0.75, transform shape}]
\begin{feynman}
\vertex(a);
\vertex[above left =1cm and 0.5cm of a] (a1){$e^{+}$};
\vertex[below left =1cm and 0.5cm of a] (a2){$e^{-}$};
\vertex[right = 1cm of a] (c);
\vertex[above right =1cm and 0.5cm of c] (b1){\({\color{black}\rm DM} \)};
\vertex[below right =1cm and 0.5cm of c] (b2){\({\color{black}\rm DM}\)};
\vertex[right = 0.75cm of c] (b3){\({\color{black}\rm\gamma}\)};
\diagram*{
(a) -- [ fermion, arrow size=0.7pt,edge label={\(\rm \)},style=black] (a1),
(a2) -- [ fermion, arrow size=0.7pt,edge label={\(\rm \)},style=black] (a),
(a) -- [ boson, edge label={\(\rm{\color{black} \gamma} \)},style=gray!75] (c),
(c) -- [plain, arrow size=0.7pt,edge label={\(\rm \)},style=red,edge label={\({\color{black}} \)}] (b1),
(b2) -- [plain, arrow size=0.7pt,style=red,edge label={\({\color{black}\rm} \)}] (c),
(b3) -- [boson ,style=red,edge label={\({\color{black}\rm} \)}] (c)};
\end{feynman}
\node at (c)[black,fill,style=black,inner sep=2pt]{};
\end{tikzpicture}
\hspace{1cm}
\begin{tikzpicture}%[baseline={(current bounding box.center)},style={scale=0.75, transform shape}]
\begin{feynman}
\vertex(a);
\vertex[above left =0.25cm and 1cm of a] (a1){$e^{+}$};
\vertex[below = 0.75cm of a](b);
\vertex[below = 0.75cm of b](c);
\vertex[below left =0.25cm and 1cm of c] (a2){$e^{-}$};
\vertex[above right =0.25cm and 1cm of b] (b1){\({\color{black}\rm DM} \)};
\vertex[below right =0.25cm and 1cm of b] (b2){\({\color{black}\rm DM}\)};
\vertex[above right = 0.25cm and 1cm of a] (d1){\({\color{black}\rm e^{+}}\)};
\vertex[below right = 0.25cm and 1cm of c] (d2){\({\color{black}\rm e^{-}}\)};
\diagram*{
(a) -- [ fermion, arrow size=0.7pt,edge label={\(\rm \)},style=black] (a1),
(a2) -- [ fermion, arrow size=0.7pt,edge label={\(\rm \)},style=black] (c),
(a) -- [ boson, edge label'={\(\rm{\color{black} \gamma/Z} \)},style=gray!75] (b),
(b) -- [ boson, edge label'={\(\rm{\color{black} \gamma/Z} \)},style=gray!75] (c),
(b2) -- [plain, arrow size=0.7pt,edge label={\(\rm \)},style=red,edge label={\({\color{black}} \)}] (b),
(b) -- [plain, arrow size=0.7pt,style=red,edge label={\({\color{black}\rm} \)}] (b1),
(d1) -- [fermion, arrow size=0.7pt,edge label={\(\rm \)},style=red,edge label={\({\color{black}} \)}] (a),
(c) -- [fermion, arrow size=0.7pt,style=red,edge label={\({\color{black}\rm} \)}] (d2)};
\end{feynman}
\node at (b)[black,fill,style=black,inner sep=2pt]{};
\end{tikzpicture}
\caption{Feynman graphs showing associated production of DM pair with mono-$\gamma$ (left) and {\tt OSE} (right) at the $e^{+} e^{-}$ collider.}
\label{fig:ldm-cd}
\end{figure}

\begin{figure}[htb!]
\centering
\begin{tikzpicture}
\begin{feynman}
\vertex(a);
\vertex[above left =0.4cm and 0.25cm of a] (a0);
\vertex[above right =0.5cm and 1cm of a0] (g){$\gamma$};
\vertex[above left =0.5cm and 0.25cm of a0] (a1){$e^{+}$};
\vertex[below left =1cm and 0.5cm of a] (a2){$e^{-}$};
\vertex[right = 1.5cm of a] (c);
\vertex[above right =1cm and 0.5cm of c] (b1){\({\color{black}\rm \nu} \)};
\vertex[below right =1cm and 0.5cm of c] (b2){\({\color{black}\rm \overline{\nu}}\)};
\diagram*{
(a) -- [ fermion, arrow size=0.7pt,edge label={\(\rm \)},style=black] (a0),
(a0) -- [ fermion, arrow size=0.7pt,edge label={\(\rm \)},style=black] (a1),
(a2) -- [ fermion, arrow size=0.7pt,edge label={\(\rm \)},style=black] (a),
(a0) -- [ boson, style=red,edge label={\({\color{black}} \)}] (g),
(a) -- [ boson, edge label={\(\rm{\color{black} Z} \)},style=gray!75] (c),
(c) -- [fermion, arrow size=0.7pt,edge label={\(\rm \)},style=red,edge label={\({\color{black}} \)}] (b1),
(b2) -- [fermion, arrow size=0.7pt,style=red,edge label={\({\color{black}\rm} \)}] (c)};
\end{feynman}
\end{tikzpicture}
\hspace{1cm}
\begin{tikzpicture}
\begin{feynman}
\vertex(a);
\vertex[above left =0.3cm and 0.5cm of a] (a0);
\vertex[above right =0.5cm and 1cm of a0] (g){$\gamma$};
\vertex[above left =0.2cm and 0.5cm of a0] (a1){$e^{+}$};
\vertex[below = 1cm of a] (c);
\vertex[below left =0.5cm and 1cm of c] (a2){$e^{-}$};
\vertex[above right =0.5cm and 1cm of a] (b1){\({\color{black}\rm \nu} \)};
\vertex[below right =0.5cm and 1cm of c] (b2){\({\color{black}\rm \overline{\nu}}\)};
\diagram*{
(a) -- [ fermion, arrow size=0.7pt,edge label={\(\rm \)},style=black] (a0),
(a0) -- [ fermion, arrow size=0.7pt,edge label={\(\rm \)},style=black] (a1),
(a2) -- [ fermion, arrow size=0.7pt,edge label={\(\rm \)},style=black] (c),
(a0) -- [ boson, style=red,edge label={\({\color{black}} \)}] (g),
(a) -- [ boson, edge label={\(\rm{\color{black} W} \)},style=gray!75] (c),
(a) -- [fermion, arrow size=0.7pt,edge label={\(\rm \)},style=red,edge label={\({\color{black}} \)}] (b1),
(b2) -- [fermion, arrow size=0.7pt,style=red,edge label={\({\color{black}\rm} \)}] (c)};
\end{feynman}
\end{tikzpicture}
\hspace{1cm}
\begin{tikzpicture}
\begin{feynman}
\vertex(a);
\vertex[above left =0.1cm and 1cm of a] (a1){$e^{+}$};
\vertex[below = 0.9cm of a] (c1);
\vertex[right = 1cm of c1] (d) {$\gamma$};
\vertex[below = 0.9cm of c1] (c);
\vertex[below left =0.1cm and 1cm of c] (a2){$e^{-}$};
\vertex[above right =0.1cm and 1cm of a] (b1){\({\color{black}\rm \nu} \)};
\vertex[below right =0.1cm and 1cm of c] (b2){\({\color{black}\rm \overline{\nu}}\)};
\diagram*{
(a) -- [ fermion, arrow size=0.7pt,edge label={\(\rm \)},style=black] (a1),
(a2) -- [ fermion, arrow size=0.7pt,edge label={\(\rm \)},style=black] (c),
(c1) -- [ boson, style=red,edge label={\({\color{black}} \)}] (d),
(a) -- [ boson, edge label={\(\rm{\color{black} W} \)},style=gray!75] (c1),
(c1) -- [ boson, edge label={\(\rm{\color{black} W} \)},style=gray!75] (c),
(a) -- [fermion, arrow size=0.7pt,edge label={\(\rm \)},style=red,edge label={\({\color{black}} \)}] (b1),
(b2) -- [fermion, arrow size=0.7pt,style=red,edge label={\({\color{black}\rm} \)}] (c)};
\end{feynman}
\end{tikzpicture}
\caption{Feynman graphs for relevant SM backgrounds contributing to mono-$\gamma$+ missing energy ($\slashed{E}$) for DM signal at the the $e^+e^-$ colliders.}
\label{fig:sm-bkg}
\end{figure}
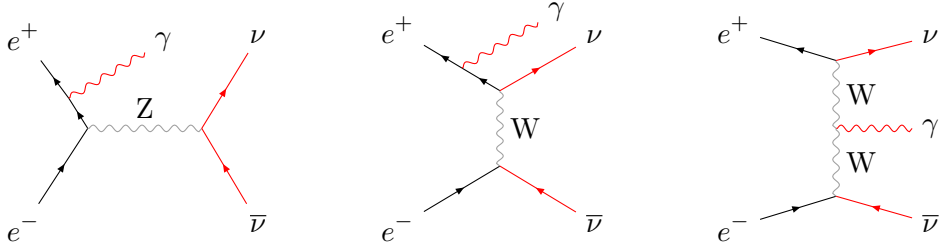

%%%%%%%%%%%%%%%%%%%%%%%%%%%%%%%%
Identifying mono-$\gamma$ events is one of the most effective methods for searching DM at the colliders. Dominant non-interfering SM neutrino background contribution for mono-$\gamma$ signal arises from $Z$-mediated $s$-channel and $W$-mediated $t$-channel diagrams as shown in Fig.~\ref{fig:sm-bkg}. Most analyses of DM searches through mono-$\gamma$ final state signal have focused on considering ISR photon. This approach often results in signal distributions that closely resemble those of the background for certain collider variables. As a consequence, it becomes difficult to segregate the signal from the SM background. On the other hand, the operators [cf. Eq.~\eqref{eq:eft-ops}] we consider, lead to the associated production of a photon with the pair production of DM particles, resulting in the natural mono-$\gamma$ signal [cf. Fig.~\ref{fig:ldm-cd}]. This signal produces distinct event distributions (compared to the SM background) for certain collider variables, enabling the introduction of cuts that can effectively segregate the SM background from the signal to some extent.
%%%%%%%%%%%%%%%%%%%%%%%
%%%%%%%%%%%%
%%%%%%%%%%%%%%%%%%%%%%%%%%

The model implementation is done using {\tt FeynRules} \cite{Alloul:2013bka} and we use {\tt MG5\_aMC} \cite{Alwall:2011uj} to generate signal and background Monte-Carlo (MC) events. The event files are then showered using {\tt Pythia8} \cite{Sjostrand:2014zea} and detector simulation is executed through {\tt Delphes3} \cite{deFavereau:2013fsa}. We limit the phase space during event generation by implementing the transverse momentum of a photon, $p_T^\gamma > 10$ GeV, and pseudorapidity, $|\eta_{\gamma}| \le 2.5$ cuts. Additionally, we only consider events with a single photon and no detected leptons or jets in the final state. Below we define the kinematical variables that we have utilized for signal-background analysis:
\begin{itemize}
\item {\bf Missing transverse energy ($\slashed{E}_T$)}: Missing Transverse Energy (MET) can be estimated from the momentum imbalance in the transverse direction associated with visible particles, providing an indication of missing particles that are not registered by the detector. MET is defined as
\begin{equation}
\slashed{E}_T= - \sqrt{\left(\sum p_x\right)^2+\left(\sum p_y\right)^2}\,,
\end{equation}
where $p_{x,\,y}$ are the 3-momentum along $x$ and $y$ direction respectively, and the sum runs over all the visible particles registered in the detector.
\item {\bf Missing energy ($\slashed{E}$)}: The energy carried away by the missing particles is defined as missing energy (ME) which can be identified at a lepton collider, given the information of the CM energy of the process. ME is expressed as
\begin{equation}
\slashed{E}=\sqrt{s}-\sum_{i} E_i\,,
\end{equation}
where $i$'s are the possible visible particles detected in the detector.
\item {\bf Pseudorapidity ($\eta_{\gamma}$)}: We define pseudorapidity from the knowledge of photon emerging angle as
\begin{equation}
\eta_{\gamma}=-\ln\left[\tan \left(\frac{\theta_{\gamma}}{2} \right)\right]\,.
\end{equation}
\item {\bf Azimuthal angle ($\phi_{\gamma}$)}: The azimuthal angle for the photon is defined as
\begin{equation}
\phi_{\gamma}= \tan^{-1} \left(\frac{p_{y}}{p_{x}}\right),
\end{equation}
where, $p_{x}$ and $p_{y}$ are the $x$ and $y$ component of the 4-momenta of the photon.
\end{itemize}
\begin{figure}[htb!]
\centering
\includegraphics[width=0.475\linewidth]{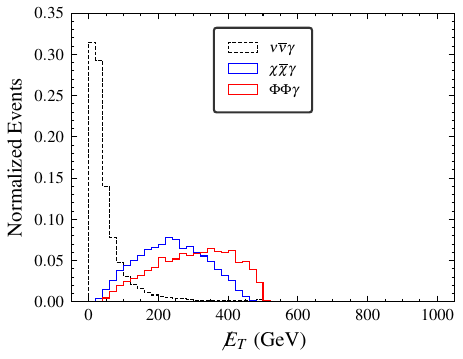}
\includegraphics[width=0.475\linewidth]{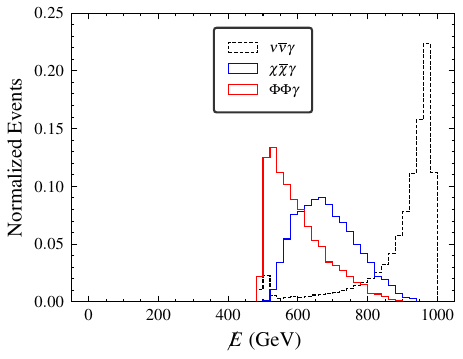}
\includegraphics[width=0.475\linewidth]{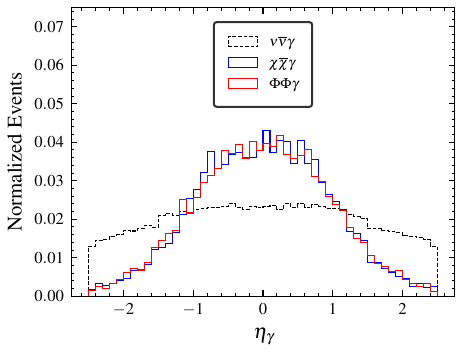}
\includegraphics[width=0.475\linewidth]{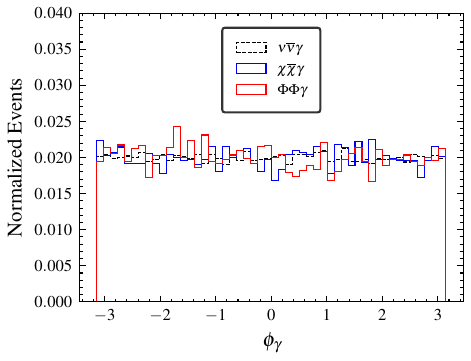}
\caption{Event distribution for mono-$\gamma$ plus missing energy signal and the SM background at the ILC with $\sqrt{s}$ = 1 TeV. We choose $m_{\chi} = 100$ GeV, $\Lambda_{\rm NP} = 1.2$ TeV for the fermionic DM and $m_{\Phi} = 500$ MeV, $\Lambda_{\rm NP} = 2$ TeV for the scalar DM.}
\label{fig:dist1}
\end{figure}
%%%%%%%%%%%%%%%%%%%%%%%%

In Fig.~\ref{fig:dist1} we illustrate the signal-background event distribution for the different kinematical variables defined above. The ISR photon from the SM neutrino background typically exhibits low $p^{\gamma}_T$, unlike the signal photon, which is produced in association with the DM pair. This distinct feature of the MET distribution enables clear differentiation between the DM signal and the SM background, as shown in the top left panel of Fig.~\ref{fig:dist1}. The missing energy distribution of the SM background displays a double peak behavior, as can be noted in the top right plot of Fig.~\ref{fig:dist1}. The peak at $\slashed{E}\sim 1$ TeV is attributed to the dominant $W$-mediated $t$-channel diagram, while another sub-dominant peak at the tail of the distribution $\slashed{E}\sim 500$ GeV, is due to the $Z$-mediated $s$-channel diagram. The peak position at the tail can be identified using the expression
\begin{equation}
\slashed{E}=\frac{\sqrt{s}}{2}\left(1+\frac{m_Z^2}{s}\right),
\end{equation}
where $m_Z$ is the $Z$-boson mass. On the other hand, for the signal distribution where the DM pair is produced along with a photon, it is clear that the ME distribution peaks near 500 GeV (for $\sqrt{s}$ = 1 TeV) and then exhibits a continuously declining trend. Thus, applying a well-considered cut on ME significantly eliminates a large portion of the SM neutrino background. Finally, applying an absolute pseudorapidity cut of $ |\eta_{\gamma}| < 1.0 $ effectively refines the removal of non-transverse backgrounds (bottom left panel of Fig.~\ref{fig:dist1}). As we see from the bottom right panel, $\phi_{\gamma}$ inefficient to discriminate the signal from the background.

The total polarized cross-section considering partial beam polarization ($-100\%<P_{e^{\pm}}<100\%$) is given by~\cite{Fujii:2018mli,Bambade:2019fyw,Barman:2021hhg}
\begin{equation}
\sigma(P_{e^+},\,P_{e^-}) = \sum_{\lambda_{e^+}= \pm1 }\,\sum_{ \lambda_{e^-}= \pm1} \frac{(1+ \lambda_{e^-} P_{e^-})\,(1+ \lambda_{e^+} P_{e^+})}4\, \sigma_{\lambda_{e^-}, \lambda_{e^+} } \,,
\end{equation}
where $\lambda_i=-1(+1)$ is the left (right)-handed helicity of electron/positron and  $P_{e^{\pm}}$ is positron (electron) beam polarization. In the SM, left-handed leptons have stronger couplings with gauge bosons compared to right-handed ones. Consequently, employing a right-polarized electron beam and a left-polarized positron beam will be advantageous in reducing the SM background. Following the ILC Snowmass report \cite{ILCInternationalDevelopmentTeam:2022izu}, we choose $\{P_{e^{+}}, P_{e^{-}}\} = \{-20 \%, +80 \%\}$ combination which provides about six-fold suppression of the SM background while enhancing the signal approximately by 16\% as evident from Table~\ref{tab:cut1} and \ref{tab:cut2}. The signal significance is presented as \cite{Cowan:2010js} 
\begin{equation}
\mathcal{S} = \sqrt{2 \left[(S + B) \log \left(1 + \frac{S}{B} \right) - S \right]}\,,
\end{equation}
where $S$ and $B$ are the signal and background events, respectively. By carefully selecting beam polarization and applying subsequent cuts on the aforementioned collider variables, the signal and background events, along with the signal significance, are listed in Table~\ref{tab:cut1} (Table~\ref{tab:cut2}) for fermion (scalar) DM for the benchmark: $m_{\chi} (m_{\Phi}) = 100$ GeV (500 MeV) and $\Lambda_{\rm NP} = 1.2$ TeV (2 TeV) at $\mathfrak{L}_{\text{int}}$ = 8 $\rm{ab^{-1}}$. We observe about $99 \%$ reduction in the SM neutrino background after employing all the relevant cuts, while retaining around $50\%$ of the signal in each case.
%%%%%%%%%%%%%%%%%%%%%%
\begin{table}[h!]
\centering
\begin{tabular}{|c|c|c|c|c|c|c|}
\hline
\multirow{2}{*}{Cuts} & \multicolumn{3}{c|}{$\{P_{e^{+}}, P_{e^{-}}\} = \{0 \%, 0 \%\}$} & \multicolumn{3}{c|}{$\{P_{e^{+}}, P_{e^{-}}\} = \{-20 \%, +80 \%\}$} \\ \cline{2-7}
& $\chi \overline{\chi} \gamma$ & $\nu \overline{\nu} \gamma$ & Significance & $\chi \overline{\chi} \gamma$ & $\nu \overline{\nu} \gamma$ & Significance \\ \hline
Basic cuts & 2371 & 18061101 & 0.56 & 2747 & 3455860 & 1.48 \\
$\slashed{E}_{T} > 190$ GeV & 1595 & 790250 & 1.79 & 1808 & 447846 & 2.70 \\
$\slashed{E} \in (510, 750)$ GeV & 1462 & 406112 & 2.29 & 1659 & 142321 & 4.39 \\
$|\eta_{\gamma}| < 1$ & 1291 & 272872 & 2.47 & 1462 & 79403 & 5.17 \\
\hline
\end{tabular}
\caption{Cutflow for signal and background events for mono-$\gamma$ signal at the ILC with $\sqrt{s}$ = 1 TeV and $\mathfrak{L}_{\text{int}}$ = 8 $\rm{ab^{-1}}$ for unpolarized ($\{P_{e^{+}}, P_{e^{-}}\} = \{0 \%, 0 \%\}$) and polarized ($\{P_{e^{+}}, P_{e^{-}}\} = \{+20 \%, -80 \%\}$) cases. Here, we consider fermionic DM with mass $m_{\chi} = 100$ GeV and $\Lambda_{\rm NP} = 1.2$ TeV.}
\label{tab:cut1}
\end{table}
%%%%%%%%%%%%%%%%%%%%%%%
\begin{table}[h!]
\centering
\begin{tabular}{|c|c|c|c|c|c|c|}
\hline
\multirow{2}{*}{Cuts} & \multicolumn{3}{c|}{$\{P_{e^{+}}, P_{e^{-}}\} = \{0 \%, 0 \%\}$} & \multicolumn{3}{c|}{$\{P_{e^{+}}, P_{e^{-}}\} = \{-20 \%, +80 \%\}$} \\ \cline{2-7}
& $\Phi\Phi\gamma$ & $\nu \overline{\nu} \gamma$ & Significance & $\Phi\Phi\gamma$ & $\nu \overline{\nu} \gamma$ & Significance \\ \hline
Basic cuts & 2945 & 18061101 & 0.69 & 3425 & 3455860 & 1.84 \\
$\slashed{E}_{T} > 190$ GeV & 2406 & 790250 & 2.70 & 2802 & 447846 & 4.18 \\
$\slashed{E} \in (510, 750)$ GeV & 2134 & 406112 & 3.34 & 2509 & 142321 & 6.63 \\
$|\eta_{\gamma}| < 1$ & 1762 & 272872 & 3.37 & 2075 & 79403 & 7.33 \\
\hline
\end{tabular}
\caption{Same as Table~\ref{tab:cut1}, but for scalar DM with $m_{\Phi}=500$ MeV and $\Lambda_{\rm NP} = 2$ TeV.}
\label{tab:cut2}
\end{table}
%%%%%%%%%%%%%%%%%%%%%%
\subsection{Opposite-sign electron signal}
\begin{figure}[htb!]
\centering
\includegraphics[width=0.475\linewidth]{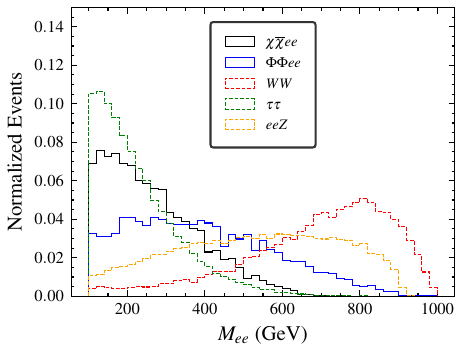}
\includegraphics[width=0.475\linewidth]{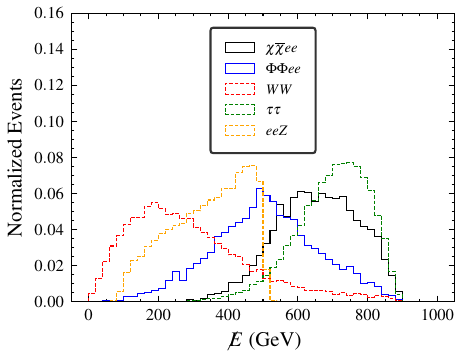}
\includegraphics[width=0.475\linewidth]{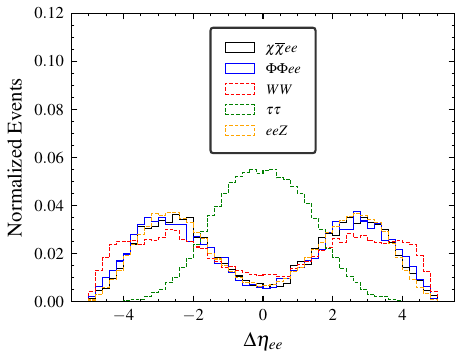}
\includegraphics[width=0.475\linewidth]{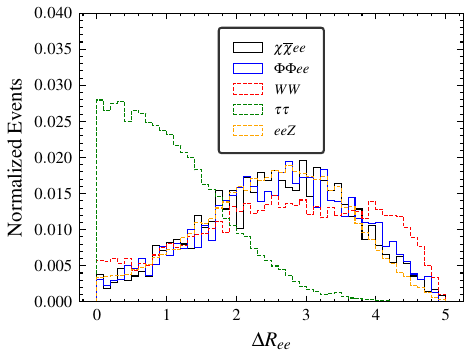}
\caption{Event distribution of dilepton plus missing energy signal and corresponding SM background for $\sqrt{s}$ = 1 TeV. We choose $m_{\chi} = 100$ GeV, $\Lambda_{\rm NP} = 1.2$ TeV for the fermionic DM and $m_{\Phi} = 100$ MeV, $\Lambda_{\rm NP} = 2$ TeV for the scalar DM.}
\label{fig:dist2}
\end{figure}

\begin{figure}[htb!]
\centering
\includegraphics[scale=0.22]{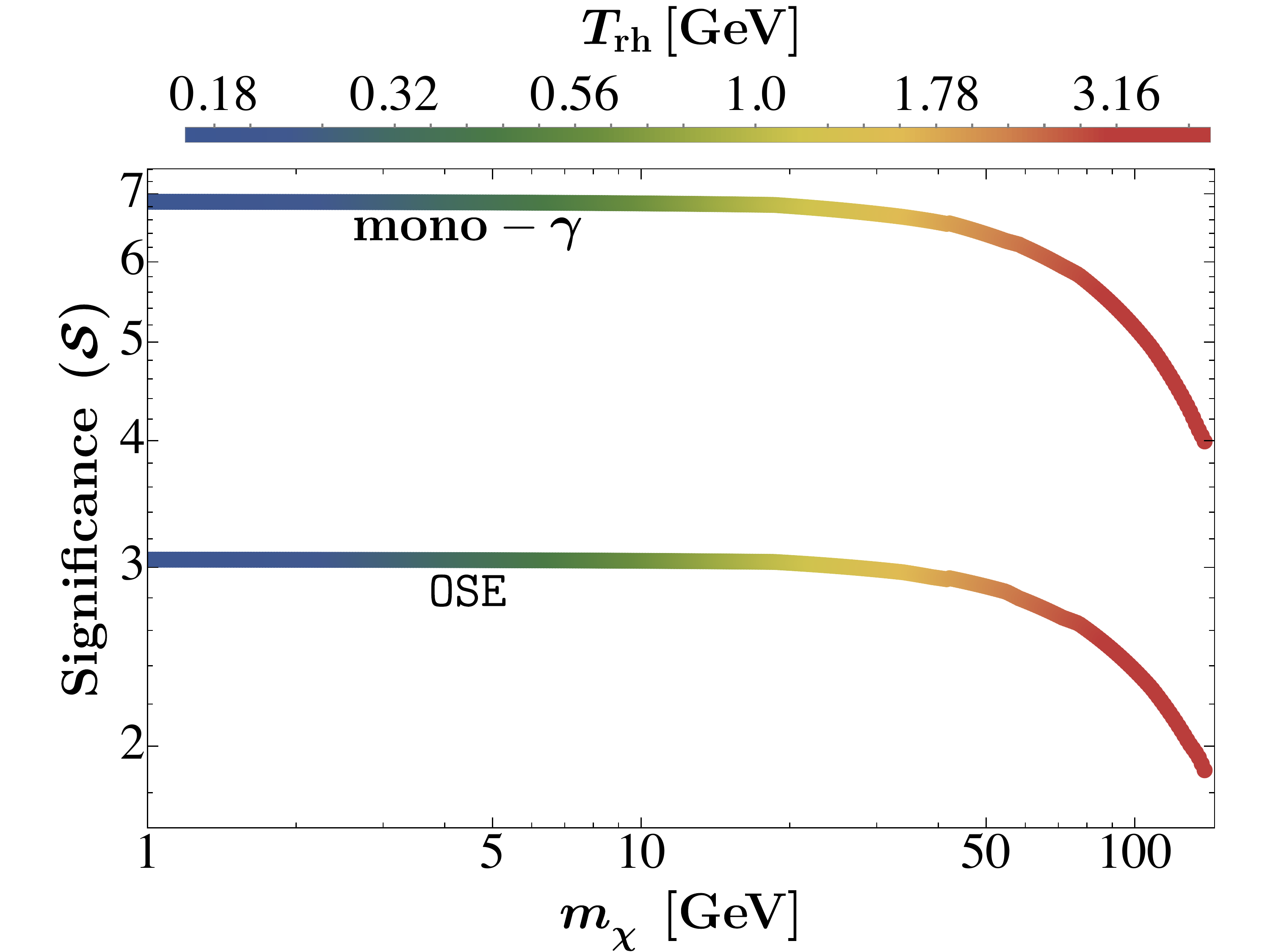}
\caption{Signal significance as a function of DM mass for the fermionic DM. We choose $n=6$, $\Lambda_{\rm NP}=1.2 ~\rm TeV$, $\sqrt{s}=\rm 1~TeV$ and $\mathfrak{L}_{\text{int}}=8~\rm{ab^{-1}}$. The polarization combinations are $\{P_{e^{+}}, P_{e^{-}}\} = \{-20 \%, +80 \%\}~(\{+20 \%, +80 \%\})$ for mono-$\gamma$ ({\tt OSE}) signal.  The bar legend denotes the corresponding reheating temperature required to satisfy the observed DM abundance.}
\label{fig:colldier-scan}
\end{figure}

%%%%%%%%%%%%%%%%%%%%%%
Another possible collider signature can arise from opposite sign electron ({\tt OSE}) channel, where DM is produced through the fusion of neutral vector bosons (VBF) radiated from the initial state electrons, as shown in Fig.~\ref{fig:ldm-cd}. Charged VBF (involving fusion of $W$ boson) can also produce DM pair in association with SM neutrinos resulting in no visible final state particles\footnote{The {\tt OSE} signal can also appear from $Z$ decay to electrons arising from ``natural mono-$Z$'' possibilities similar to the ``natural mono-$\gamma$'' scenario discussed in previous section, however, this signal is subdominant and is subjected to large background suppression.}. Hence, in this section, we will focus only on the neutral VBF production mode. For the {\tt OSE} final state signal, we strictly select events with two electrons, excluding any events with detected photons, jets, and muons. The dominating SM background processes for {\tt OSE} final state are $W^{+}W^{-}$ and $\tau^{+} \tau^{-}$ subsequent decay to electrons and neutrinos. A significant background arises from $e^{+} e^{-} Z$ production, where the $Z$ subsequently decays to neutrinos. Another important background contribution comes from $\nu \overline{\nu} Z$ production, where the $Z$ decays to electron final states. Some kinematic variables relevant to the {\tt OSE} signal are:
%%%%%%%%%%%%%%%%%%
\begin{itemize}
\item {\bf Invariant di-electron mass ($M_{ee}$)}: This is defined as
\begin{equation}
M_{ee} = \sqrt{\left(p_{e^{+}} + p_{e^{-}}\right)^{2}}\,,
\end{equation}
where $p_{e^{+}}$ ($p_{e^{-}}$) is the 4-momenta of the detected positron (electron). Invariant mass distributions peak at resonances and can be used as an important discriminating variable when the signal or background process is associated with a resonant production of a heavy particle.
\item {\bf Difference in pseudorapidity between the electron pair ($\Delta \eta_{ee}$)}: The definition is as follows:
\begin{equation}
\Delta \eta_{ee} = (\eta_{e^{+}} - \eta_{e^{-}})\,,
\end{equation}
where $\eta_{e^{+}}$ ($\eta_{e^{-}}$) is the pseudorapidity of the positron (electron).
\item {\bf Distance between the electron pair ($\Delta R_{ee}$)}: This is defined in the $\left(\eta, \phi\right)$ plane as
\begin{equation}
\Delta R_{ee} = \sqrt{\left(\Delta \eta_{ee}\right)^{2} + \left(\Delta \phi_{ee}\right)^{2}}\,,
\end{equation}
where, $\Delta \phi_{ee}$ is the difference in the azimuthal angles of the electron-positron.
\end{itemize}
%%%%%%%%%%%%%%%
In order to eliminate electrons produced from $Z$ decay, we pre-select events with a $M_{ee} > 100$ GeV. This efficiently removes the $Z-$pole and thereby discarding the mammoth background contributions from $\nu \overline{\nu} Z$ completely (also removing the ``natural mono-$Z$'' DM production in the process). The kinematic distributions of the signal and corresponding backgrounds are shown in Fig.~\ref{fig:dist2}.
%%%%%%%%%%%%%%%%%%%%%%%%%%%
%%%%%%%%%%%%%%%%%%%%%%%%%
%%%%%%%%%%%%%%%%%%%%%%%%%%%%%%%
\begin{figure}[htb!]
\centering
\includegraphics[width=0.475\linewidth]{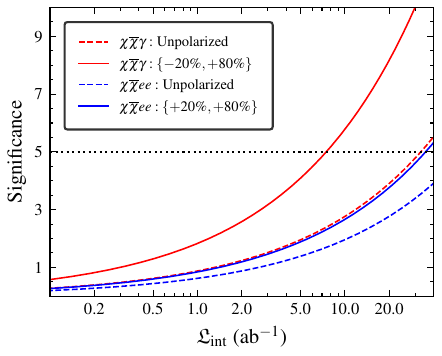}
\includegraphics[width=0.475\linewidth]{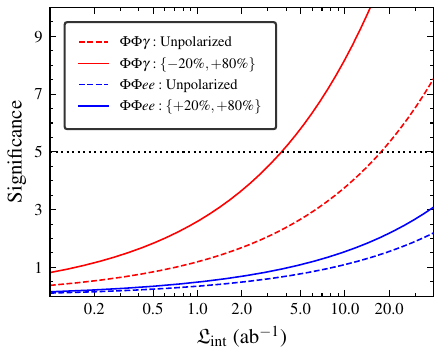}
\caption{Signal significance as a function of integrated luminosity for a different combination of beam polarizations $\{P_{e^{+}}, P_{e^{-}}\}$ in case of fermionic DM (left) and scalar DM, (right). We choose $m_{\chi} = 100$ GeV and $\Lambda_{\rm NP} = 1.2$ TeV for the fermion DM and $m_{\Phi} = 500$ MeV and $\Lambda_{\rm NP} = 2$ TeV, with $\sqrt{s}$ = 1 TeV. The dotted black line corresponds to $5\sigma$ significance.} 
\label{fig:lum-sig}
\end{figure}
The missing energy distribution (top right of Fig.~\ref{fig:dist2}) peaks opposite for the signal compared to the dominant background $W^{+}W^{-}$ and $e^{+}e^{-}Z$, hence a cut demanding $\slashed{E} > 550$ GeV, significantly kills these background processes. Similarly, considering the $\Delta \eta_{ee}$ distribution (top left of Fig.~\ref{fig:dist2}), we observe that the $\tau^{+} \tau^{-}$ background is peaked at the center whereas the signal is bi-peaked symmetrically away from the center. A judicious cut on $\Delta \eta_{ee}$ ($|\Delta \eta_{ee}| > 1$) is applied to remove the central region of the distribution which eliminates most of the $\tau \tau$ background \footnote{Similar signal-background discrimination can be obtained if $\Delta{R}_{ee}$ is used instead.}. Following subsequent cuts, the signal and background events, along with the signal significance, are listed in Table~\ref{tab:cut3} (Table~\ref{tab:cut4}) for fermion (scalar) DM for the benchmark: $m_{\chi} (m_{\Phi}) = 100$ GeV (500 MeV) and $\Lambda_{\rm NP} = 1.2$ TeV (2 TeV) at $\mathfrak{L}_{\text{int}}$ = 8 $\rm{ab^{-1}}$ for the unpolarized as well as the optimal polarization choice of $\{P_{e^{+}}, P_{e^{-}}\} = \{+20 \%, +80 \%\}$. This polarization combination suppresses the total background by a factor of 1.5 while enhancing the signal marginally. We observe a $\sim 96 \%$ reduction in the SM backgrounds while retaining around $55\%$ of the signal in each case after the implementation of subsequent cuts on the above-mentioned kinematical collider variables. 
%%%%%%%%%%%%%%%%%%%%%%
\begin{table}[h!]
\centering
\begin{tabular}{|c|c|c|c|c|c|c|}
\hline
\multirow{2}{*}{Cuts} & \multicolumn{3}{c|}{$\{P_{e^{+}}, P_{e^{-}}\} = \{0 \%, 0 \%\}$} & \multicolumn{3}{c|}{$\{P_{e^{+}}, P_{e^{-}}\} = \{+20 \%, +80 \%\}$} \\ \cline{2-7}
& $\chi \overline{\chi} ee$ & Backgrounds & Significance & $\chi \overline{\chi} ee$ & Backgrounds & Significance \\ \hline
Basic cuts & 222 & 131699 & 0.61 & 235 & 80923 & 0.83 \\
$\slashed{E} > 550$ GeV & 178 & 15931 & 1.41 & 189 & 10255 & 1.86 \\
$|\Delta \eta_{ee}| > 1$ & 121 & 4793 & 1.75 & 128 & 2876 & 2.38 \\
\hline
\end{tabular}
\caption{Cutflow for signal and background events for \texttt{OSE} signal at the ILC with $\sqrt{s}$ = 1 TeV and $\mathfrak{L}_{\text{int}}$ = 8 $\rm{ab^{-1}}$ for unpolarized ($\{P_{e^{+}}, P_{e^{-}}\} = \{0 \%, 0 \%\}$) and polarized ($\{P_{e^{+}}, P_{e^{-}}\} = \{+20 \%, +80 \%\}$) cases. Here we consider fermionic DM with $m_{\chi} = 100$ GeV and $\Lambda_{\rm NP} = 1.2$ TeV.}
\label{tab:cut3}
\end{table}
%%%%%%%%%%%%%%%%%%%%%%%%%%
\begin{table}[h!]
\centering
\begin{tabular}{|c|c|c|c|c|c|c|}
\hline
\multirow{2}{*}{Cuts} & \multicolumn{3}{c|}{$\{P_{e^{+}}, P_{e^{-}}\} = \{0 \%, 0 \%\}$} & \multicolumn{3}{c|}{$\{P_{e^{+}}, P_{e^{-}}\} = \{+20 \%, +80 \%\}$} \\ \cline{2-7}
& $\Phi\Phi ee$ & Backgrounds & Significance & $\Phi\Phi ee$ & Background & Significance \\ \hline
Basic cuts & 273 & 131699 & 0.75 & 279 & 80923 & 0.98 \\
$\slashed{E} > 550$ GeV & 103 & 15931 & 0.81 & 110 & 10255 & 1.09 \\
$|\Delta \eta_{ee}| > 1$ & 68 & 4793 & 0.98 & 74 & 2876 & 1.38 \\
\hline
\end{tabular}
\caption{Same as Table~\ref{tab:cut3} but for scalar DM with mass $m_{\Phi}=500$ MeV and $\Lambda_{\rm NP} = 2$ TeV.}
\label{tab:cut4}
\end{table}
%%%%%%%%%%%%%%%%%%%%%%%%%%%%
The signal significance as a function of the integrated luminosity for different final states and combination of polarizations are illustrated in Fig.~\ref{fig:lum-sig}. We note that the scalar DM is expected to achieve $5\sigma$ signal significance at a lower integrated luminosity compared to the fermionic candidate. This is because of the fact that the scalar DM emerges from a dimension-six operator in contrast to the fermionic DM, which results from a dimension-seven operator, resulting in an additional suppression if the mass and NP scale benchmarks are left identical. This points toward the possibility of detection of the scalar candidate at a much lower luminosity corresponding to the early runs of the ILC, whereas the fermionic DM might be probed only at the high luminosity runs. Similarly, the variation of signal significance with the DM mass $\mdm$ for the relic density allowed points (for $n=6$) is shown in Fig.~\ref{fig:colldier-scan} for both the spin-0 and spin-1/2 DM candidates. The significance drops with the increase in DM mass, simply because of the non-availability of final state phase space that results in smaller production cross-section at the collider. Here we also denote the corresponding $\Trh$ that is required to satisfy the observed DM abundance. We remind the readers once again that in this case only the bosonic reheating scenario is valid as in that case $\lNP>\Tmax$ is satisfied in order to achieve the observed abundance for DM mass of our interest. This enables us to establish a one-to-one correspondence between the DM mass and the reheat temperature, that are controlled by the reheating dynamics, with the signal significance at the ILC.
%%%%%%%%%%%%%%%%%%%%%%%%%%
%%%%%%%%%%%%%%%%%%%%%%
%%%%%%%%%%%%%%%%%%%%%%%
\section{Conclusion}
\label{sec:concl}
%%%%%%%%%%%%%%%%%%%%%
In this work, we have explored the ability of lepton colliders to shed light on the pre-BBN cosmology. To establish a one-to-one correspondence between collider and cosmological observables, we focused on dark matter (DM) genesis via UV freeze-in, which requires non-renormalizable interactions between DM and the visible sector, i.e., the Standard Model (SM). We considered DM production during the era of reheating, facilitated by DM-SM operators with mass dimensions six and seven—where the former pertains to spin-0 DM, and the latter to spin-1/2 DM. There is a potential difficulty in probing UV freezing in DM at collider, as the effective NP scale requires to be quite large compared to the DM mass. 
This evidently makes the signal cross-section much smaller compared to weak interaction cross-section. Therefore, the operators that can produce a distinguishable signal from the SM background at collider, need 
to produce a missing energy or MET distribution characteristically different from that of the background. This is what happens when DM operators couple to SM field strength 
tensors, the resulting mono-photon signal can be segregated from the SM contamination with judicious missing energy and MET cuts at future lepton colliders. It is a bit difficult to
do the same exercise in context of LHC, as the background events are huge and the separability is not that clean.

As a concrete model for the pre-BBN cosmology, we assumed reheating occurred through the perturbative decay of the inflaton into either a pair of SM-like bosons or fermions, with the inflaton $\phi$ oscillating at the bottom of a monomial potential $V(\phi)\propto\phi^n$ during this period. Such monomial potentials are well-motivated from CMB measurements of inflationary observables, for example, the tensor-to-scalar ratio, spectral index or the amplitude of the tensor power spectrum. A detailed analysis of the reheating epoch allows 
us to explore beyond reheating temperature as $\Trh<\Tmax$.

For significant collider signals to emerge over the SM background, the scale of NP needs to be around $\sim\mathcal{O}$(TeV). We found that in this scenario, the observed DM abundance can be achieved only through bosonic reheating, with a reheating temperature $\Trh\sim\mathcal{O}$(MeV), such that the effective description of the DM-SM interaction remains valid during reheating [cf. Fig.~\ref{fig:mdm-trh}]. {\color{black} The scalar DM operator connecting to SM via field strength tensor gets highly suppressed from indirect search bounds from two photon annihilation channel. We study collider signal excess that can be achieved via mono photon signal for both scalar and fermionic DM in relic density and indirect search limited regions.} This implies that any signal excess detected, for instance, in the mono-photon channel (or opposite-sign electron channel) at the lepton collider (e.g., the International Linear Collider), could potentially indicate reheating via a bosonic channel and a low (MeV) reheating temperature [cf. Fig.~\ref{fig:colldier-scan}]. In conclusion, this analysis not only provides a method to probe freeze-in at colliders but also offers insight into the earliest epoch of the Universe, demonstrating the power of particle colliders in investigating the pre-BBN era.
%%%%%%%%%%%%%%%%%%
\appendix
%%%%%%%%%%%%%%%%
\section{Relevant annihilation cross-sections}
\label{sec:cs}
%%%%%%%%%%%%%%%
Here we report the annihilation cross-sections of a pair of SM states into a pair of spin-0 DM of mass $m_\Phi$
\begin{align}
\sigma_{\gamma\gamma\to \Phi\Phi}\,&=\,\frac{s}{4 \pi  \lNP^4}\sqrt{1-4\,m_\Phi^2/s}\,,\\
\sigma_{VV\to \Phi\Phi}\,&=\frac{s}{9\pi\Lambda ^4}\sqrt{\frac{1-4\,m_\Phi^2/s}{1-4\,m_V^2/s}}\left(1-4\,m_V^2/s+6\, (m_V^2/s)^2\right)\,,
\end{align}
a pair of spin-1/2 DM of mass $m_\chi$
\begin{align}
\sigma_{\gamma\gamma\to \chi\overline{\chi}}\,&=\,\frac{s^2}{4 \pi  \lNP^6} (1-4\,m_\chi^2/s)^{3/2}\,,\\
\sigma_{VV\to \chi\overline{\chi}}\,&=\frac{s^2}{9 \pi  \Lambda ^6}\dfrac{(1-4\,m_\chi^2/s)^{3/2}}{(1-4 \,m_V^2/s)^{1/2}} \left(1-4\,m_V^2/s+6(m_V^2/s)^2\right)\,,
\end{align}
and into a pair of spin-1 DM of mass $m_X$
\begin{align}
\sigma_{\gamma\gamma\to XX}\,&=\,\frac{s^3}{240\pi\,\lNP^4\,m_X^4}\sqrt{1-4\,m_X^2/s}\left(17-36\,m_X^2/s+292\,(m_X^2/s)^2\right)\,,\\
\nonumber\sigma_{VV\to XX}\,&=\,\frac{s^3}{540 \pi  \lNP^4 m_X^4}\sqrt{\frac{s-4 m_X^2}{s-4 m_V^2}}\bigg[17-36 (m_X^2/s)+292 (m_X^2/s)^2-2 (m_X^2/s) 
\\&
\left(568 (m_X^2/s)^2 
-24(m_X^2/s)+23\right)+4(m_X^2/s)^2\left(688(m_X^2/s)^2-84(m_X^2/s)+23\right)\bigg]\,,
\end{align}
where $V=\{W,\,Z\}$ represents the SM massive gauge bosons. The reaction density then reads
\begin{align}
\gamma_{\text{SM}\,\text{SM}\to\text{DM}\,\text{DM}} &=\int\prod_{i=1}^4 d\Pi_i \left(2\pi\right)^4 \delta^{(4)}\biggl(p_a+p_b-p_1-p_2\biggr)\,f_a{^\text{eq}}f_b{^\text{eq}}\left|\mathcal{M}_{a,b\to1,2}\right|^2
\nonumber\\&
=\frac{T}{32\pi^4}\,g_a g_b\,\int_{s_\text{min}}^{\lNP^2} ds\,\frac{\biggl[\bigl(s-m_a^2-m_b^2\bigr)^2-4m_a^2 m_b^2\biggr]}{\sqrt{s}}\,\sigma\left(s\right)_{a,b\to1,2}\,K_1\left(\frac{\sqrt{s}}{T}\right)\label{eq:gam-ann}\,, \end{align}    
with $a,b~(1,2)$ as the incoming (outgoing) states and $g_{a,b}$ are corresponding degrees of freedom. Here $f_i{^\text{eq}}\approx g_i\exp^{-E_i/T}$ is the Maxwell-Boltzmann distribution. The Lorentz invariant 2-body phase space is denoted by: $d\Pi_i=\frac{d^3p_i}{\left(2\pi\right)^3 2E_i}$. The amplitude squared (summed over final and averaged over initial states) is denoted by $\left|\mathcal{M}_{a,b\to1,2}\right|^2$ for a particular 2-to-2 scattering process. The lower limit of the integration over $s$ is $s_\text{min}=\text{max}\left[\left(m_a+m_b\right)^2,\left(m_1+m_2\right)^2\right]$.
%%%%%%%%%%%%%%%%%%%%%%%%%
\bibliographystyle{JHEP}
\bibliography{ldm}
\end{document}